\documentclass{aa}  

\usepackage{natbib}
\bibpunct{(}{)}{;}{a}{}{,}
\usepackage{graphicx}
\usepackage{txfonts}
\usepackage{hyperref}
\usepackage{multicol}
\usepackage{subfig}
\usepackage{hyperref}
\usepackage{appendix}
\usepackage{multirow}
\usepackage{footnote}

\DeclareUnicodeCharacter{2212}{-}
\hypersetup{draft}  
\begin{document} 

\title{The evolution of galaxy scaling relations in clusters at 0.5<z<1.5 
\thanks{Based on observations made with the Gran Telescopio Canarias (GTC), installed at the Spanish Observatorio del Roque de los Muchachos of the Instituto de Astrofísica de Canarias, in the island of La Palma. PI: Helmut Dannerbauer. Program's IDs: 122-GTC70/17A and 137-GTC118/18A.}}
 
\author{J. M. P\'erez-Mart\'inez\inst{1,2,3}
           \and B. Ziegler\inst{1}
           \and H. Dannerbauer\inst{2,3}
           \and A. B\"ohm\inst{1}
           \and M. Verdugo\inst{1}
           \and A. I. D\'iaz\inst{4}
           \and C. Hoyos\inst{4}
          }
      
\institute{Department of Astrophysics, University of Vienna, A-1180 Vienna, Austria. \email{jm.perez@univie.ac.at}
\and Instituto de Astrof\'isica de Canarias (IAC), E-38205, La Laguna, Tenerife, Spain.  
\and Universidad de La Laguna, Dpto. Astrof\'isica, E-38206, La Laguna, Tenerife, Spain
\and Departamento de F\'isica Te\'orica and CIAFF, Universidad Aut\'onoma de Madrid, E-28049 Madrid, Spain
}

\date{}

\abstract
{}
{We present new gas kinematic observations with the OSIRIS instrument at the GTC for galaxies in the Cl1604 cluster system at $z\sim0.9$. These observations together with a collection of other cluster samples at different epochs analyzed by our group are used to study the evolution of the Tully-Fisher, velocity-size and stellar mass-angular momentum relations in dense environments over cosmic time.}
{We use 2D and 3D spectroscopy to analyze the kinematics of our cluster galaxies and extract their maximum rotation velocities ($V_{max}$), which will be used as the common parameter in all scaling relations under scrutiny. We determine the structural parameters of our objects by fitting surface brightness profiles to the images of our objects, while stellar-mass values are computed via SED fitting by making use of extensive archival optical to NIR photometry. Our methods are consistently applied to all our cluster samples which make them ideal for an evolutionary comparison. }  
{Up to redshift one, our cluster samples show evolutionary trends compatible with previous observational results in the field and in accordance with semianalytical models and hydrodynamical simulations concerning the Tully-Fisher and velocity-size relations. However, we find a factor $\sim$3 drop in disk sizes and an average B-band luminosity enhancement $\langle\Delta M_B\rangle\sim2$ mag by z$\sim$1.5. We discuss the role that different cluster-specific interactions may play in producing this observational result. In addition, we find that our intermediate-to-high redshift cluster galaxies follow parallel sequences with respect to the local specific angular momentum-stellar mass relation, although displaying lower angular momentum values in comparison with field samples at similar redshifts. This can be understood by the stronger interacting nature of dense environments with respect to the field.} 
{}



\maketitle


\section{Introduction}
\label{S:Intro} 

Scaling relations are strong trends between the main physical parameters of galaxies and they are key to understand the different processes at play in galaxy evolution. In spiral galaxies, the flat part of their rotation curves provides us with a proxy, the maximum circular velocity, to trace the total mass of galaxies (including dark matter). This allows us to study the interplay between the dark and the baryonic component of galaxies by making use of some other easily observable parameters such as the stellar-mass (or luminosity) and disk size. These parameters define a three-dimensional space with the potential to describe most of the physical transformations that a galaxy experience during its lifetime. The different projections of this space yield several important scaling relations (\citealt{Koda00}) that can be also reproduced by assuming the virial equilibrium of structures and the conservation of angular momentum during the dissipational collapse of cold dark matter haloes (\citealt{Mo98}, \citealt{vandenBosch00}, \citealt{Navarro00}). Some of the simplest and yet most fundamental scaling relations for spiral galaxies are the Tully-Fisher relation (TFR) and the velocity-size relation (VSR), which were first observed by \citet{Tully77}. However, more complex parameter combinations produce other interesting relations such as the angular momentum-stellar mass (\citealt{Fall83}, \citealt{Romanowsky12}) that are key to understand the processes of morphological transformation and mass redistribution during galaxy evolution. 

The TFR connects the $V_{max}$ (taken as the rotation velocity in the flat part of the rotation curve) and the luminosity (or stellar-mass) of a spiral galaxy.  During the last decades, the field TFR has been studied in depth up to z$\sim$2 (\citealt{Tully98}, \citealt{Ziegler02}, \citealt{Kassin07}, \citealt{Puech08}, \citealt{Miller11}, \citealt{Boehm16}, \citealt{Tiley16}, \citealt{Simons16}, \citealt{Harrison17}, \citealt{Pelliccia17}, \citealt{Ubler17}). Different representations of the TFR provide information about the evolution of the galaxies' stellar populations and their stellar mass growth across cosmic time. For example, the study of the B-band TFR yields a luminosity enhancement of up to 1 mag by z$\sim$1 in the field (\citealt{Boehm16}), which is in agreement with the predicted gradual evolution towards younger stellar populations in galaxies with lookback time (\citealt{Dutton11}). 

It is expected that the stellar-mass of galaxies grows with time due to the progressive consumption of their gas reservoirs. However, the exact evolution of the stellar-mass TFR is still a matter of debate, especially at high redshift. Some authors claim a strong evolution (around -0.4 dex in $M_*$) at z$\sim$2 in comparison with the local universe (\citealt{Price16}, \citealt{Straatman17}), while others show results compatible with a mild to negligible evolution at z$\sim$1 (\citealt{Miller11}, \citealt{Contini16}, \citealt{diTeodoro16}, \citealt{Pelliccia17}). However, \cite{Tiley16} and \cite{Ubler17} find similar strong offsets between these two epochs. These conflicting results may arise from the varying morphological and kinematic selection criteria applied in each study and the difficulty to identify rotation dominated systems at high redshift. Interestingly, \citet{Tiley16} report $\Delta M_*\approx-0.4$ in the stellar-mass TFR for galaxies at z$\sim$1 displaying high rotational support ($V/\sigma>3$). However, this offset disappears when they consider their full sample of galaxies, regardless of the individual $V/\sigma$ values. Thus, the use of a common methodological frame is required when investigating the evolution of scaling relations at different redshifts.

The velocity-size relation traces the growth of disks inside evolving Navarro-Frenk-White dark matter haloes (NFW, \citealt{Navarro97}). However, this correlation is weaker than the TFR and displays a wider scatter (\citealt{Courteau07}, \citealt{Hall12}). This is partially explained due to the ambiguities in defining the size of a galaxy ($R_e$, $R_d$, or other prescriptions) at different wavelengths taking into account the evolution and distribution of the different stellar populations within the galaxy. Furthermore, the presence of a bulge component and additional selection effects (surface brightness limits) may contribute to hinder its study (\citealt{Meurer18}, \citealt{Lapi18}). Nevertheless, in the context of galaxy evolution, this scaling relation remains one of the tools to look for environmental imprints over the disks.

During the early phases of galaxy formation, the angular momentum of the collapsing dark matter haloes is transferred to the baryonic matter. This process is key to understand the early formation of disks and the distribution of baryonic matter within them. Thus, the study of the angular momentum allows us to simultaneously connect the rotation velocity, the stellar-mass, and the galaxy size into a single scaling relation: the specific angular momentum-stellar mass relation (\citealt{Fall83}), which can be influenced over time by several processes such as morphological transformations, galaxy interactions and the presence of inflows. For example, \citet{Romanowsky12} observed a decreasing trend in the specific angular momentum of galaxies with increasing bulge-to-disc ratio (see also \citealt{Fall13} and \citealt{Fall18}), linking the morphological transformation of galaxies to the redistribution of their angular momentum. 

In clusters, the baryonic and the dark component of galaxies can be influenced by cluster-specific interactions, either related with the intracluster medium (strangulation and ram-pressure stripping) or due to gravitational interactions caused by the high number density of galaxies in this environment (harassment, tidal interactions, and mergers). Up to z$\sim$1, it has been reported similar evolution in the cluster and field environments with respect to the B-band TFR (\citealt{Ziegler03}, \citealt{Jaffe11}, \citealt{Bosch2}), while others claim a mild luminosity enhancement (\citealt{Bamford05}) and larger TFR scatter in the cluster environment (\citealt{Moran07}, \citealt{Pelliccia19}). However, the VSR and the angular momentum evolution have been scarcely studied in dense environments up to date. In this work, we gather several cluster samples studied by our group in the past together with recent GTC/OSIRIS observation over the multicluster system Cl1604+4304 to consistently investigate the possible influence of the environment over kinematic scaling relations across cosmic time. The Tully-Fisher (TFR), the Velocity-Size (VSR), and the angular momentum stellar-mass relation ($j-M_*$, AMR hereafter) provide a unique way to search for signs of environmental evolution over the population of cluster galaxies at different cosmological epochs. This work is structured in the following way: $\text{Sect.}$ 2 describes the main characteristics of the cluster and field samples that we collected to study the different scaling relations. $\text{Sect.}$ 3 contains a description of the methods used to analyze the CL1604 sample, which are also applied to all our cluster samples. $\text{Sect.}$ 4 and 5 are respectively devoted to the presentation and discussion of our results while $\text{Sect.}$ 6 outlines the major conclusions of this study. Throughout this article we assume a \citet{Chabrier03} initial mass function (IMF), and adopt a flat cosmology with $\Omega_{\Lambda}$=0.7, $\Omega_{m}$=0.3, and $H_{0}$=70 km s$^{-1}$Mpc$^{-1}$. All magnitudes quoted in this paper are in the AB system.

\section{Sample Overview}
\label{S:Sample}

In this section, we describe the main characteristics of the several datasets used in this work. Our cluster sample is composed of galaxies from six distant cluster and multicluster systems studied by our group:  CL1604+4304 at z$\sim$0.9 (hereafter CL1604) which is introduced in this work, XMMUJ2235-2557 at z$\sim$1.4 (hereafter XMM2235, \citealt{JM17}), HSC-CL2329 and HSC-CL2330 at z$\sim$1.47 (hereafter HSC-protoclusters, \citealt{Boehm20}), RXJ1347-1145 at z$\sim$0.45 (hereafter RXJ1347, \citealt{JM20}) and Abell 901/902 at z$\sim$0.16 (\citealt{Bosch1, Bosch2}, hereafter the Abell-clusters). A summary of the main cluster properties ($M_{200}$, $R_{200}$, $\sigma$) of each (sub-)structure is provided in Table\,\ref{T:clusters}. We emphasize that the kinematic analysis of all the cluster samples was carried out using the same techniques that have been applied in this study (see Sect. \ref{SS:Methods}). This allows a direct comparison to test the evolution of scaling relations in dense environments at different epochs. For this same reason, we use \cite{Boehm16} as our main comparison sample in the \textit{field} for the TFR and VSR analysis. The simultaneous study of the TFR, VSR, and AMR requires the measurement of B-band luminosity, stellar-mass ($M_*$), rotation velocity ($V_{max}$) and effective radius ($R_e$) for every cluster galaxy. This information is not always available in all our samples, and thus we define a primary cluster sample made of galaxies from RXJ1347, CL1604 and XMM2235, that will be subject to our full scaling relation analysis, while the Abell and HSC clusters will only be considered as comparison cluster samples in the TFR. All the relevant parameters of the cluster primary sample can be found in the appendix (tables \ref{T:CL16}, \ref{T:RXJ} and \ref{T:XMM}).

In addition, we add field samples from other research groups to examine the evolution of the specific angular momentum relation ($j_*-M_*$, hereafter AMR) at $0<z<2.5$ (\citealt{Harrison17}, \citealt{Fall18}, \citealt{Posti18}, \citealt{Forster-Schreiber18}, \citealt{Gillman20}). This allows us to explore the environmental imprints of galaxy evolution over cosmic time. A short description of the origin and main properties of these datasets is provided in the following subsections, together with their usage and limitations in the context of the Tully-Fisher, Velocity-Size and Angular Momentum relations. However, we refer to their main publications for a more in-depth discussion of each sample's characteristics.


\begin{table*}
\centering
\caption{General properties of the clusters investigated in this study: IDs, right ascension, declination, redshift, virial mass ($M_{200}$) and size ($R_{200}$), velocity dispersion and references for these measurements.}
\begin{tabular}{ccccccccc}
\hline
\noalign{\vskip 0.1cm}
ID & RA & DEC & $z$  & $M_{200}$ & $R_{200}$ & $\sigma$ & References \\
   & (hh:mm:ss) & (dd:mm:ss) &   & $(10^{14}\,M_{\odot})$ & $(Mpc)$ & $(km/s)$ & \\
\noalign{\vskip 0.1cm}
\hline 
\hline 
\noalign{\vskip 0.2cm}
Abell 901/902      & - & - & - & - & - & - & - \\
A901a              & 09:56:27 & -09:57:22 & 0.16 & $1.3\pm0.3$ & $0.8\pm0.1$ & $880\pm30$ &   \\
A901b              & 09:55:57 & -09:59:03 & 0.16 & $1.3\pm0.3$ & $0.8\pm0.1$ & $940\pm20$ & \citealt{Heymans08}   \\
A902               & 09:56:34 & -10:10:00 & 0.17 & $0.4\pm0.2$ & $0.6\pm0.1$ & $810\pm20$ & \citealt{Weinzirl17}   \\
SW Group           & 09:55:39 & -10:10:19 & 0.17 & $0.6\pm0.2$ & $0.6\pm0.1$ & $590\pm40$ &    \\
\noalign{\vskip 0.1cm}
\hline 
\noalign{\vskip 0.2cm}
RXJ 1347        & - & - & - & - & - & - & - \\
RXJ 1347-1145      & 13:47:31 & -11:45:10 & 0.45 & $11.6\pm3.0$ & $1.9\pm0.2$ & $1160\pm100$ & \multirow{2}{*}{\citealt{Lu10}}  \\
RXJ 1347-1145      & 13:46:27 & -11:54:28 & 0.47 & $5.6\pm1.6$  & $1.2\pm0.2$ & $780\pm100$ &   \\
\noalign{\vskip 0.1cm}
\hline 
\noalign{\vskip 0.2cm}
CL1604             & - & -  & - & - & - & - & -  \\
CL1604+4304        & 16:04:22 & 43:04:56  & 0.90 & $3.3\pm1.5$ & $0.9\pm0.2$ & $720\pm130$ & \citealt{Lemaux12},  \\

CL1604+4321        & 16:04:34 & 43:21:14  & 0.92 & $1.8\pm1.6$ & $0.8\pm0.1$ & $690\pm90$ & \citealt{Wu14}   \\
\noalign{\vskip 0.1cm}
\hline 
\noalign{\vskip 0.2cm}
XMMU\,J2235.3-2557 & 22:35:21 & 25:57:40  & 1.39 & $7.3\pm1.3$ & $1.1\pm0.1$ & $1180\pm90$ & \citealt{Jee09} \\
\noalign{\vskip 0.1cm}
\hline 
\noalign{\vskip 0.2cm}
HSC-CL2329         & 23:30:05 & 00:12:36  & 1.47 & -$^a$ & - & $<400$ & \citealt{Boehm20}\\
\noalign{\vskip 0.1cm}
\hline 
\noalign{\vskip 0.2cm}
HSC-CL2330         & 23:30:22 & -00:24:00 & 1.47 & -$^a$ & - & $<400$ &  \citealt{Boehm20}\\
\noalign{\vskip 0.1cm}
\hline 
\label{T:clusters}
\end{tabular}
\tablefoot{
\tablefoottext{a}{These two proto-clusters are most probably not virialized structures yet. Thus, no velocity dispersion-based mass can be computed. See \cite{Boehm20} for more details.}
}
\end{table*}

\subsection{Cl1604 at z$\sim$0.9}
This cluster complex was first discovered by \cite{Gunn86} as two separate clusters, Cl 1604+4304 and Cl 1604+4321 at z$\sim$0.90 and 0.92 respectively. Subsequent studies made use of deep multi-band imaging and spectroscopy to unveil a much larger structure composed of several massive merging clusters and infalling groups that extends over 12 Mpc along the North-South axis (\citealt{Postman01}, \citealt{Gal04,Gal08}, \citealt{Lemaux12}, \citealt{Wu14}, \citealt{Hayashi19}). Recently, this cluster complex has been the subject of several spectrophotometric studies that have confirmed the membership of a significant number of galaxies. For example, \citealt{Crawford14, Crawford16} investigated the search for luminous compact galaxies, \citealt{Pelliccia19} investigated the kinematic evolution of cluster galaxies finding no significant differences in the TFR and lower angular momentum values in comparison to the field. In addition, \cite{Tomczak19} found that the SFR of cluster members decreases by up to 0.3 dex towards the densest regions of the cluster while \citealt{Asano20} found higher SFR for galaxies within group-like structures compared to the field and the cluster core galaxy populations.

\subsubsection{Observations}
We carried out new spectroscopic observations of cluster members in Cl1604 to study their gas kinematics by using the OSIRIS spectrograph (\citealt{Cepa00}) in MOS mode at the 10.4m Gran Telescopio de Canarias (GTC) in La Palma. Our program was split into two observing runs executed during August 2017 and June 2018 (PI: Helmut Dannerbauer. Program IDs: 122-GTC70/17A and 137-GTC118/18A respectively) under average seeing conditions of $0.8"$, airmass value of 1.3 and no moonlight contamination (i.e. dark time). We targeted two of the main structures of the cluster complex, Cl1604+4304, and Cl1604+4321, with a single mask each for a total of 16h of exposure time split between the two fields. The observations were divided into $\sim$1h observing blocks (OB) made of two on-source sub-exposures of 1420 seconds each plus overheads. We discarded one OB that was taken with seeing equal to $1.1"$. The total on-source time is 7.1h for targets in Cl1604+4321 and 4.8h for targets in Cl1604+4304. Our targets were selected from the previous works of \cite{Lemaux10, Lemaux12} and \cite{Crawford14, Crawford16}, where these objects were classified as star-forming galaxies according to their measured [OII] fluxes and blue colors.  

We extract the gas kinematics from the [OII] 3727$\AA$ emission line, which lies around 7100$\AA$ at z$\sim$0.9, with an instrumental resolution of $ \sigma_{ins} \approx 50$ km/s and a slit width of $0.9"$. To achieve this, we used the OSIRIS high-resolution grism R2500R, which covers the wavelength range $5200-7600\AA$. This configuration yielded a spectral resolution of R$\sim$2500 at the central wavelength with an average dispersion of 1 \AA/pix and an image scale of 0.25"/pix. We utilized tilted slits aligned to the apparent major axis of the targets to minimize geometrical distortions. The tilt angles $\theta$ were limited to $\left |{\theta}\right |$ < 45º to ensure a robust sky subtraction and wavelength calibration. The spectroscopic data reduction was carried out using the OSIRIS-GTCMOS pipeline (\citealt{Gomez-Gonzalez16}). The main reduction steps were bias subtraction, flat field normalization, wavelength calibration, and sky subtraction. Finally, we co-add the 2D-spectra exposures using an IRAF sigma-clipping algorithm that performs a bad pixel and cosmic ray rejection. We find no overlap between the targets that entered our kinematic analysis and the recently published work by \cite{Pelliccia19} in the same field.  

In addition to our spectroscopic campaign, we make use of the abundant complimentary archival imaging data in this field, including the Hyper Suprime-Cam Subaru Strategic Program (HSC-SSP, \citealt{Aihara18a,Aihara19}) data in five bands (g, r, i, z, y), the UKIDS survey (\citealt{Lawrence07}) in the near-infrared (J and Ks), and archival spaced-based observations with HST/ACS (F606W and F814W) and Spitzer/IRAC (3.6 and 4.5 $\mu m$). The exposure times and seeing of the retrieved co-added mosaic images are shown in Table\,\ref{T:imaging}. The coordinates, redshifts, and general properties of our final galaxy sample from this cluster are summarized in the Appendix.

\begin{table}
\caption{Summary of the photometric bands available for Cl1604}
\centering
\begin{tabular}{llcccc}
\hline
\noalign{\vskip 0.1cm}
Source &  Filter   & Exp. Time &  FWHM \\ 
          &           & (s)       &   (")   \\ \hline 
\noalign{\vskip 0.1cm}
HSC-SSP       & g             &  600  &  0.8 \\ 
\ldots        & r             &  600  &  0.8 \\
\ldots        & i             &  960  &  0.6 \\ 
\ldots        & z             &  1200 &  0.5 \\  
\ldots        & y             &  960  &  0.5\\
HST/ACS       & F606W         &  1998 &  0.1\\  
\ldots        & F814W         &  1998 &  0.1\\
Spitzer/IRAC  & 3.6$\mu m$    &  1152 &  2.0\\
\ldots        & 4.5$\mu m$    &  1152 &  2.0\\ \hline

\end{tabular}
\label{T:imaging}
\end{table}

\begin{figure*}
    \begin{multicols}{3}
      \includegraphics[width=\linewidth]{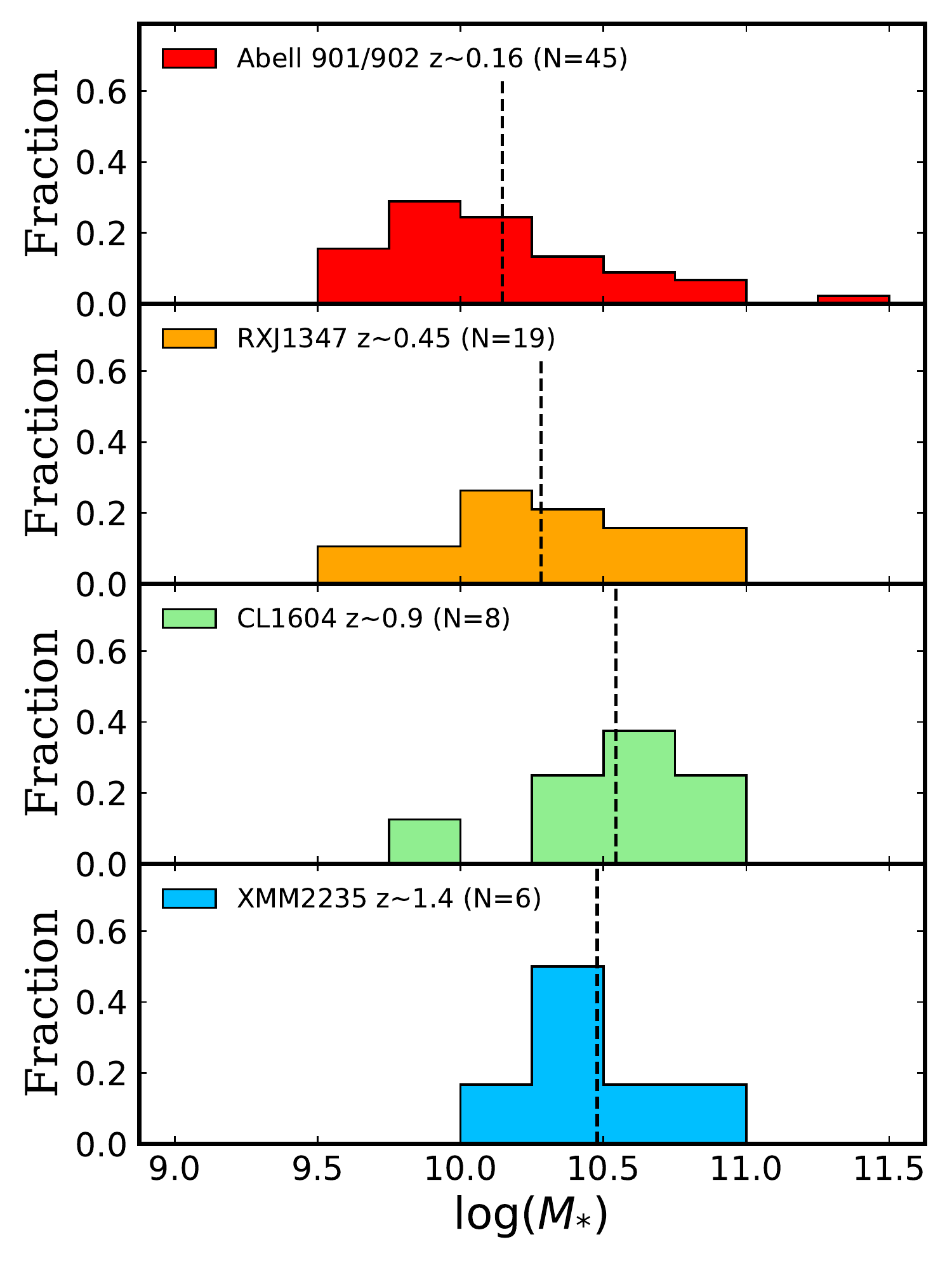}\par 
      \includegraphics[width=\linewidth]{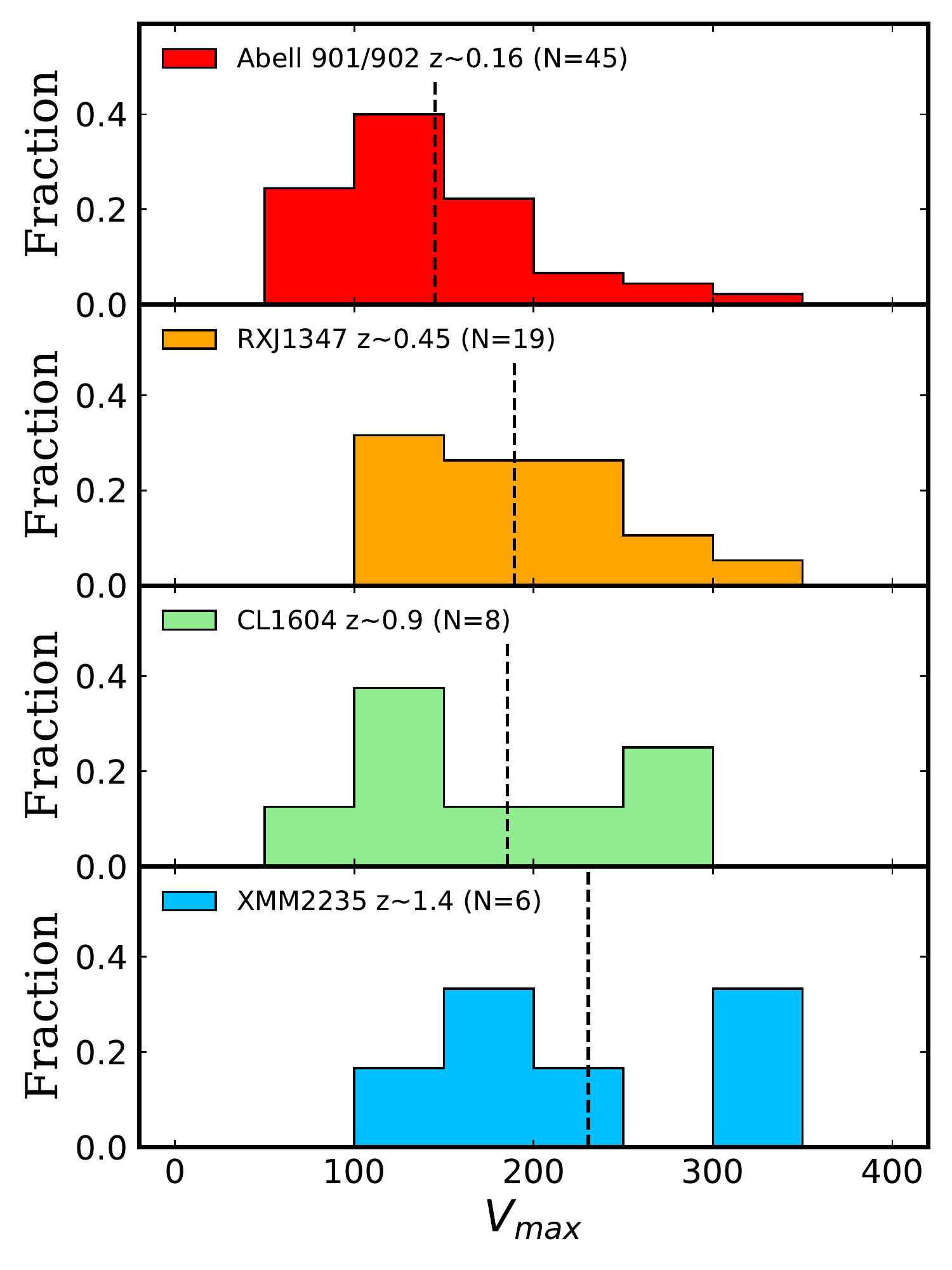}\par 
      \includegraphics[width=\linewidth]{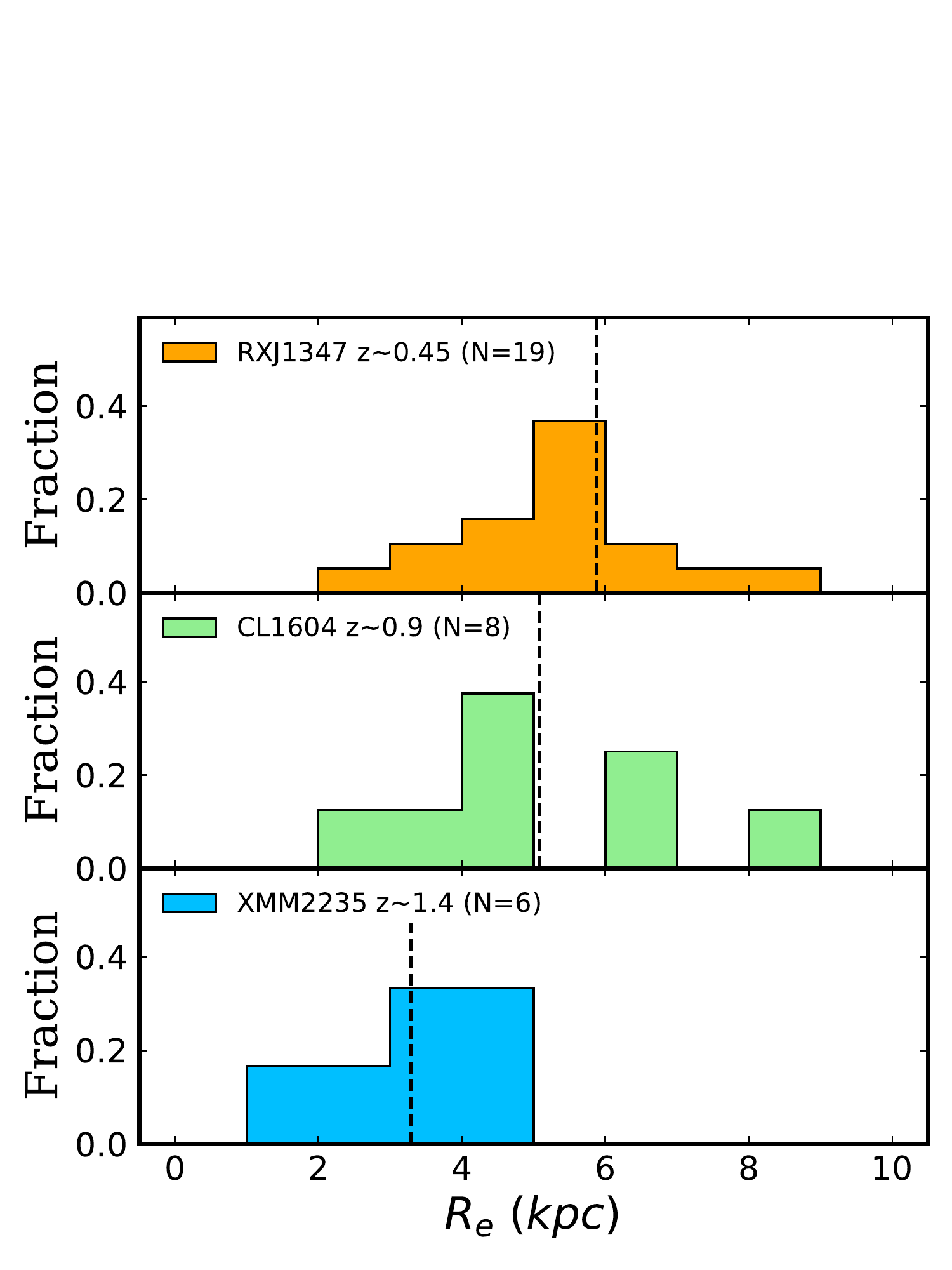}\par 
      \end{multicols}
      \caption{Distribution and mean values of the cluster samples according to the main parameters studied in this work: $M_*$, $V_{max}$, and $R_e$. The dashed lines depict the mean values for each parameter and sample. For the Abell 901/902 no effective radii are publicly available.}
         \label{F:Hist_clusters}
      \end{figure*}

\begin{figure*}
    \begin{multicols}{3}
      \includegraphics[width=\linewidth]{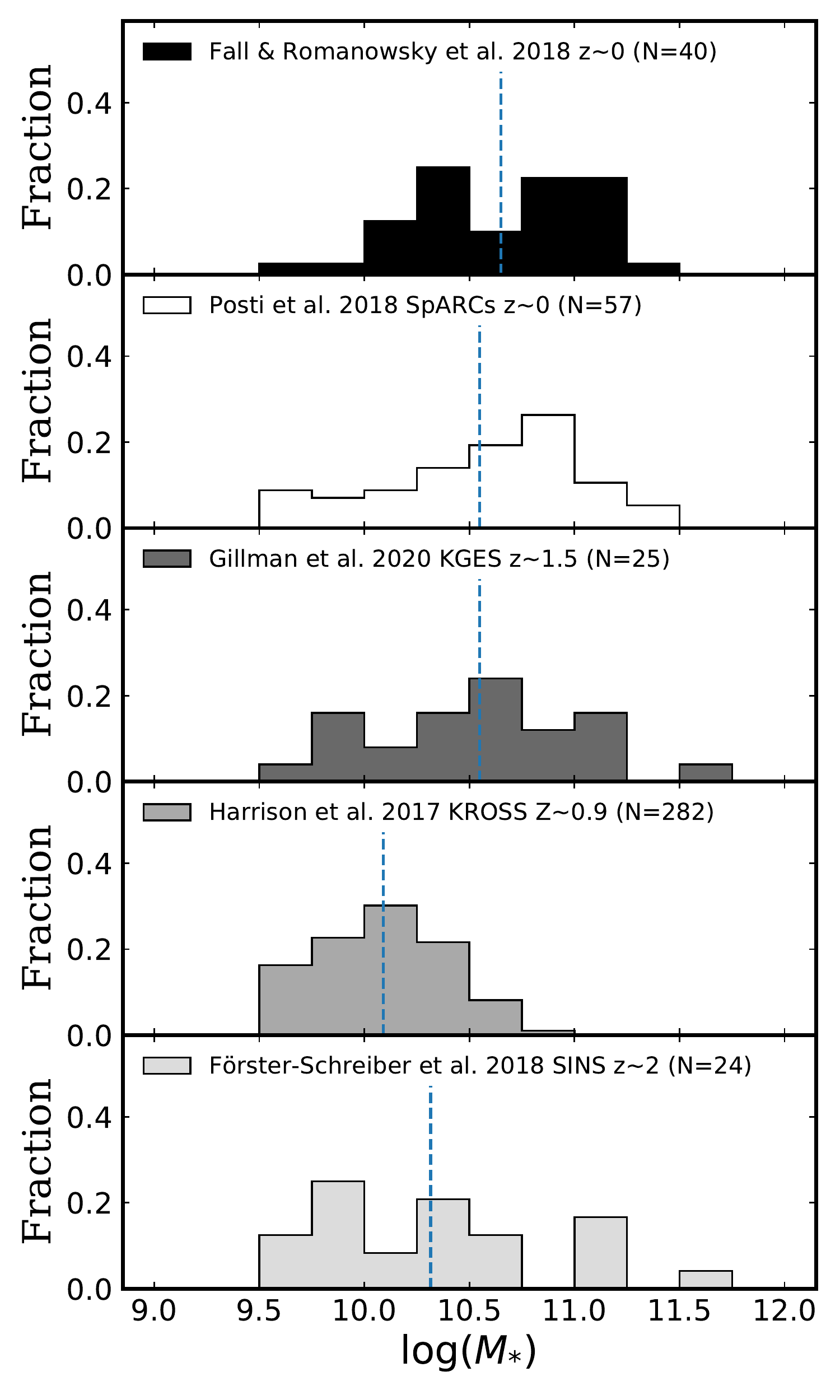}\par 
      \includegraphics[width=\linewidth]{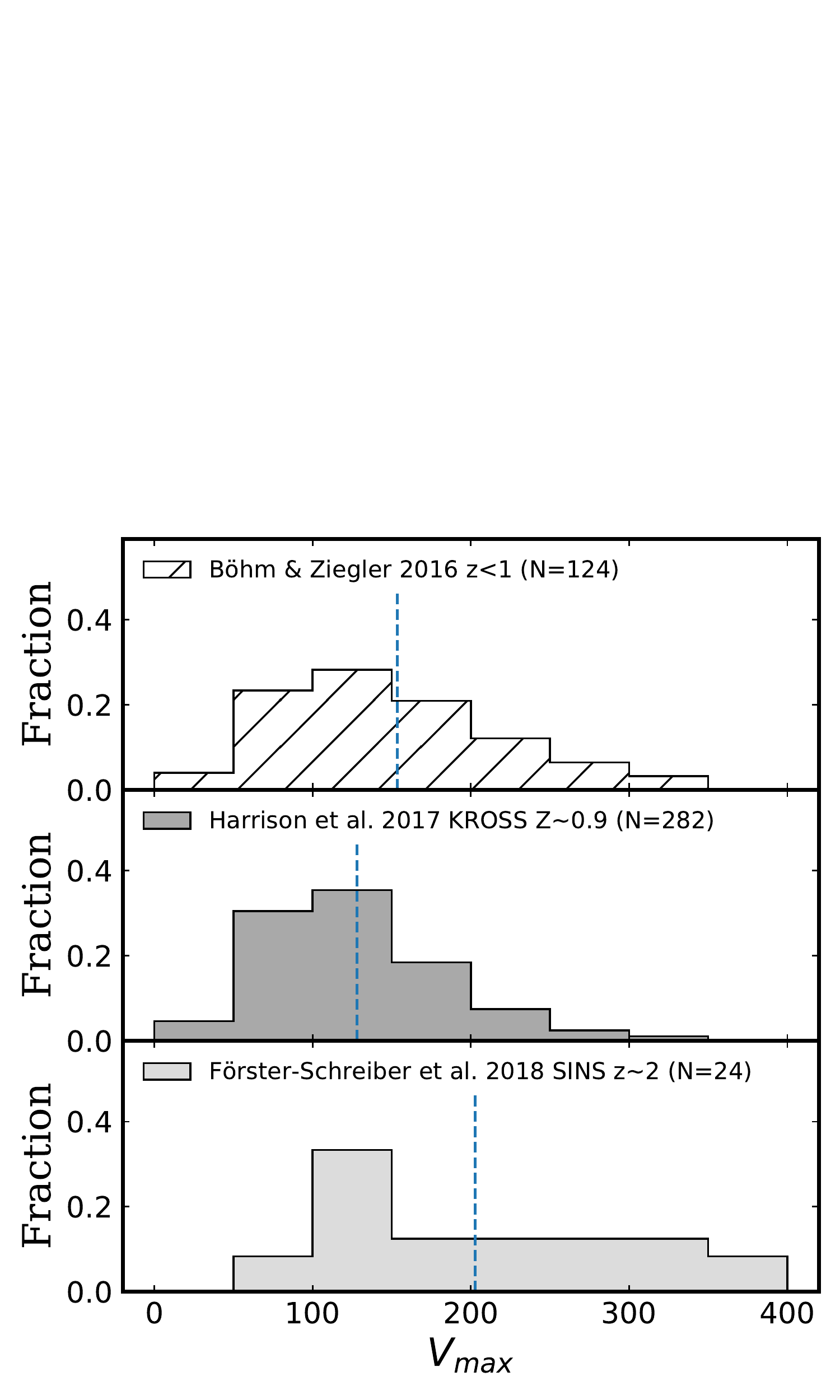}\par 
      \includegraphics[width=\linewidth]{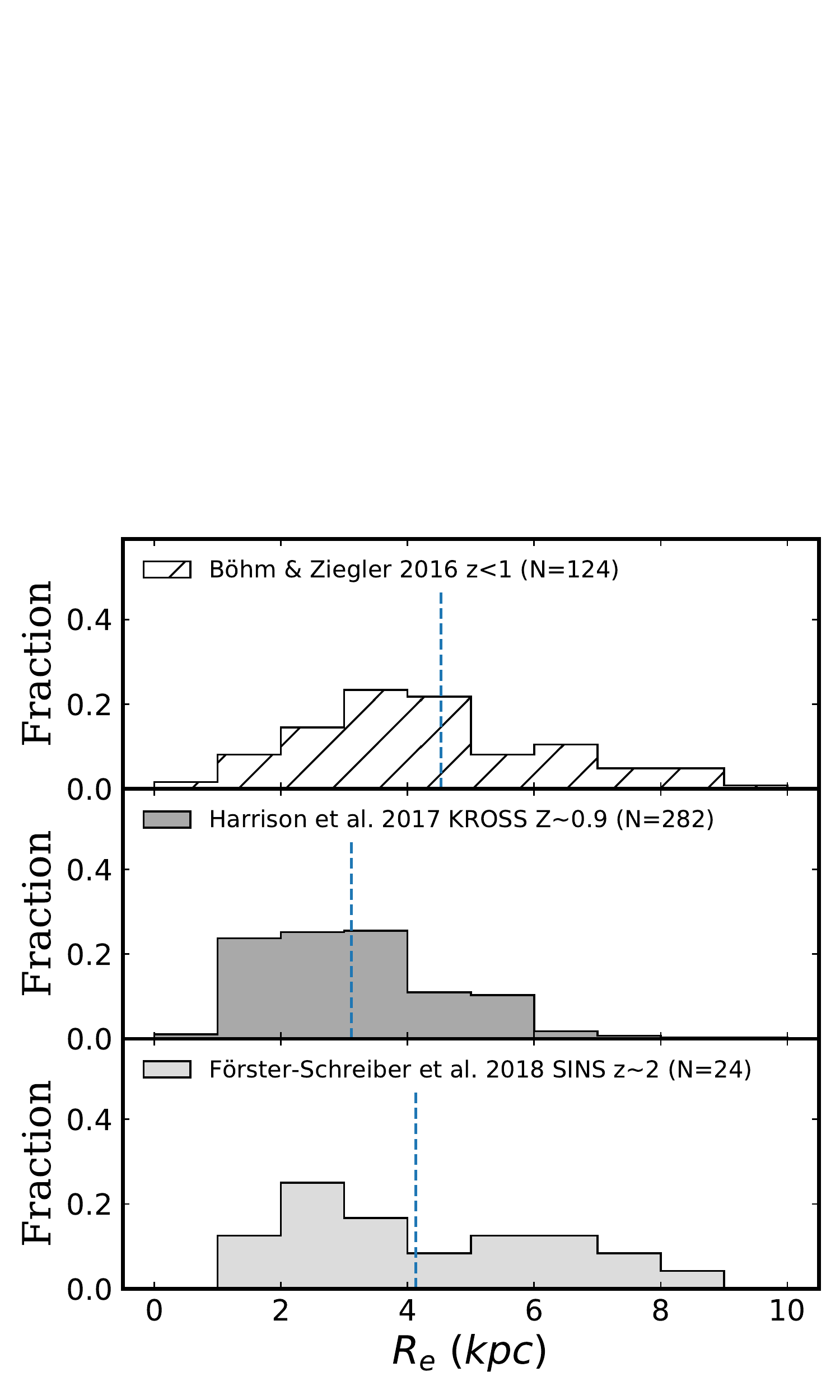}\par 
      \end{multicols}
      \caption{Distribution and mean values of the field samples according to the main parameters studied in this work: $M_*$, $V_{max}$, and $R_e$. The dashed lines depict the mean values for each parameter and sample. No effective radii and rotation velocity are publicly available for the samples of \cite{Fall18}, \cite{Posti18} and \cite{Gillman20}, while there are no stellar mass values available for \cite{Boehm16}.}
         \label{F:Hist_field}
      \end{figure*}

\subsection{Abell 901/902 at z$\sim$0.16}

This is a multicluster system composed of four main sub-structures that were intensely studied by the STAGES collaboration (\citealt{Gray09}) over the last decade. The system as a whole is not yet virialized and thus, it is an interesting laboratory to investigate the interplay between galaxy evolution and environment at low redshift. The main cluster parameters ($M_{200}$, $R_{200}$, $\sigma$) for each subsystem are provided in Table \ref{T:clusters}, and are based on the weak lensing analysis by \citealt{Heymans08} and the spectroscopic mapping of the cluster complex shown in \citealt{Weinzirl17}. 

In addition, the Abell 901/902 field has deep and extensive photometry in the UV (Galex, \citealt{Gray09}), optical (COMBO-17 survey, \citealt{Wolf03}), near-infrared (Spitzer 24$\mu m$, \citealt{Bell07}) and X-ray (XMM-Newton, \citealt{Gilmour07}) wavelength range. Out of the four main sub-structures that embody Abell 901/902, only the two most massive (A901a and A901b) show significant X-ray emission, albeit this may come from a very bright AGN close to the cluster center in A901a (\citealt{Gilmour07}). The evolution of the cluster members' colors, morphologies, and star-formation activity have been studied over the years finding redder stellar populations, lower star-formation rates and hints of interactions within the cluster population of galaxies (\citealt{Wolf09}, \citealt{Gallazzi09}). More recently, the emission-line OMEGA-OSIRIS survey provided similar results (e.g. \citealt{Chies15}, \citealt{Bruno17}, \citealt{Weinzirl17}, \citealt{Roman-Oliveira19}). Our group carried out a kinematic analysis on a sub-sample of cluster galaxies using VIMOS/VLT slit spectra taken with the HR-blue grism ($R\sim2000$) to search for indications of ram-pressure stripping (\citealt{Bosch1}) and study the slope and scatter of the Tully-Fisher relation (\citealt{Bosch2}). Galaxies within three times the velocity dispersion (3$\sigma$) of each substructure are considered to be cluster members (\citealt{Bosch1}). We retrieve a sample of 45 cluster galaxies from \citealt{Bosch2}) characterized by their high-quality rotation curves, their disc morphology, and stellar masses above $\log(M_*)=9.5$. This sample will be used in Sect. \ref{SS:Results} to explore the evolution of the TFR between clusters at different redshifts. The distribution of main parameters ($M_*$ and $V_{max}$) as well as their mean properties can be found in Fig. \ref{F:Hist_clusters} and Table \ref{T:Hist_samples}. However, the sizes and exact coordinates of these galaxies within the Abell 901/902 cluster complex are not provided in the publications aforementioned and thus we exclude this sample in the analysis that requires them.

\subsection{RXJ1347 at z$\sim$0.45}

RXJ1347 is a large-scale cluster complex composed of two merging clusters and up to 30 additional infalling groups at z$\sim$0.45 that extend diagonally across the field for about 20 Mpc (\citealt{Verdugo12}). This structure is one of the most massive and X-ray luminous clusters known (\citealt{Schindler95}). The presence of shocked gas around two very bright galaxies close to the cluster core suggests that RXJ1347 is undergoing a major merger (\citealt{Verdugo12}). Our group investigated the internal gas kinematics and star formation activity of a sample of 50 star-forming galaxies in RXJ1347 (\citealt{JM20}). Most of these objects form part of a previous medium-resolution spectroscopic campaign carried out by our group which allowed us to confirm their cluster membership and star-forming nature. The objects in our sample were distributed around the two major overdensities of the structure, the core of the central cluster (RXJ1347) and another galaxy concentration towards the south-east coinciding with the cluster LCDCS 0825 (\citealt{Gonzalez01}). The main physical parameters of these two clusters ($M_{200}$, $R_{200}$, $\sigma$, see \citealt{Lu10}) are summarized in Table \ref{T:clusters}.  

Our observations with HR-orange grism of VLT/VIMOS (R$\sim$2500) yielded 19 regular rotating objects with no significant signs of interactions in their rotation curve according to the asymmetry index criterion developed by \cite{Dale01}. In this work, we use this sample of galaxies to study the TFR, VSR, and AMR in clusters at intermediate redshift. The methodology used to obtain the main physical parameters (rest-frame magnitudes, stellar masses, sizes, and rotation velocities) is identical to the one followed to study the CL1604 sample, and described in detail in Section\,\ref{SS:Methods}. The mean properties of this cluster sample, as well as its distribution, can be found in Fig. \ref{F:Hist_clusters} and Table \ref{T:Hist_samples}.

\begin{table*}
\centering
\caption{General properties of the cluster and field samples: Cluster IDs, redshift, number of objects, mean logarithmic stellar-mass, maximum rotation velocity and effective radius and their errors.
}
\begin{tabular}{ccccccccc}
\hline
\noalign{\vskip 0.1cm} 
Cluster ID & z & N & $\overline{\log(M_{*})}$ & $\overline{V_{max}}$ & $\overline{R_{e}}$  \\
           &   &   &               & $(km/s)$  & $(kpc)$ \\
\noalign{\vskip 0.1cm}
\hline 
\hline 
\noalign{\vskip 0.2cm}
Abell 901/902    & 0.16 & 45 & $10.1\pm0.1$ & $145\pm9$ & - \\
RXJ1347         & 0.45 & 19 & $10.3\pm0.1$ & $189\pm14$ & $5.9\pm0.5$ \\
CL1604          & 0.91 & 8  & $10.5\pm0.1$ & $185\pm25$ & $5.1\pm0.6$ \\
XMM2235         & 1.39 & 6  & $10.5\pm0.1$ & $231\pm37$ & $3.3\pm0.5$ \\  
\noalign{\vskip 0.1cm}
\hline 
\noalign{\vskip 0.1cm}
\cite{Fall18}    & 0   & 40  & $10.6\pm0.1$ & -          & - \\
SpARCs              & 0   & 57  & $10.5\pm0.1$ & -          & - \\
KROSS               & 0.9 & 282 & $10.1\pm0.1$ & $128\pm3$ & $3.1\pm0.1$ \\
\cite{Boehm16}    & <1 & 124  & -            & $154\pm7$ & $4.5\pm0.2$ \\
KGES                & 1.5   & 25  & $10.5\pm0.1$ & -          & - \\
SINS                & 2.0   & 24  & $10.3\pm0.1$ & $203\pm19$ & $4.1\pm0.4$ \\  
\noalign{\vskip 0.1cm}
\hline 
\label{T:Hist_samples}
\end{tabular}
\end{table*}

\subsection{XMM2235 at z$\sim$1.4}

XMM2235 is one of the most massive and X-ray emitting virialized clusters found at $z>1$ (\citealt{Mullis05}, \citealt{Rosati09}, \citealt{Jee09}). Even though high-redshift clusters tend to be dominated by blue star-forming galaxies even in their cores (Butcher-Oemler effect, \citealt{Butcher78}), XMM2235 has a tight red sequence and a prominent BCG in its center, indicating that this cluster may be in an advanced evolutionary stage in terms of formation of its stellar populations and the assembly of its mass (\citealt{Lidman08}, \citealt{Strazzullo10}). This is also confirmed by the high $M_{200}$ and $\sigma$ values computed for this cluster via weak lensing (\citealt{Jee09}, see Table \ref{T:clusters}).

Our group carried out slit spectroscopic observations with VLT/FORS2 (R$\sim$1400) to study the gas kinematics of 27 galaxies within the cluster environment. Our target selection was based on the previous spectro-photometric campaigns carried out by \cite{Rosati09} and \cite{Strazzullo10}, as well as in the $H_{\alpha}$ narrow-band survey conducted in this field by \cite{Grutzbauch}. We successfully recovered regular rotation curves for 6 cluster members and study different scaling relations such as the Tully-Fisher and the Velocity-Size relation \citealt{JM17}. Our sample is composed of relatively compact massive objects with stellar-mass and size mean values equivalent to $\log M_*=10.5\pm0.1$ and $R_e=3.3\pm0.5$ kpc respectively (see Fig. \ref{F:TFR} and Table \ref{T:Hist_samples}). Further details about the analyses of this sample can be found in \citealt{JM17}. 


\subsection{HSC-Cl2329 and HSC-Cl2330 at z$\sim$1.47}

These clusters were identified as strong [OII] overdensities exploiting the narrow-band filter NB921 in the HSC-SSP 16deg$^2$ emission-line survey (\citealt{Hayashi18a}) and not by X-ray observations or red-sequence detection. They are dominated by star-forming galaxies and are more typical progenitors of today's regular population of clusters. Therefore, these HSC clusters offer a window to test the properties of galaxies during the cluster assembling process. Our group carried up 3D-spectroscopy with KMOS (R$\sim$4000) at VLT for both structures, confirming the membership of 34 objects and extracting regular velocity fields for 14 objects, which were used to explore the B-band Tully-Fisher relation at this redshift (\citealt{Boehm20}). Based on the low velocity dispersion and the large projected size of the proto-clusters, the authors argue that these structures are not yet virialized. However, the authors could not compute stellar-masses due to the lack of sufficient photometric bands covering the rest-frame redder part of the spectral range. Similarly, the PSF size from the HSC images matches the expected effective radius of galaxies at this redshift, which makes any attempt to determine the galaxy size unreliable. Thus, we restrict the use of this sample to the analysis of the cluster B-band TFR at different epochs. For a more detailed description of the cluster detection and target properties we refer to \cite{Hayashi18a} and \cite{Boehm20}.


\subsection{The field comparison samples}
 
In this section, we briefly introduce the main properties of the field comparison samples used in our analysis. Our primary comparison sample is composed of 124 field galaxies at z<1 from \cite{Boehm16}. These galaxies were selected from the FORS Deep Field (\citealt{Heidt03}) and the William Herschel Deep Field (\citealt{Metcalfe01}). Reliable rotation velocities, B-band absolute magnitudes, and galaxy sizes were computed for these objects following the same methods applied to our cluster sample, which are described in Sect. \ref{SS:Methods}. We will use this sample in the context of the evolution of the B-band Tully-Fisher and Velocity-Size relation whereas we exclude it from the M$_*$-TFR and the angular momentum analyses due to the lack of reliable stellar-mass values.

We also selected several kinematic studies with reliable values of angular momentum as comparison samples at different redshifts. The first sample is taken from \cite{Fall13, Fall18}. 
In these papers, the authors study the specific angular momentum of galaxies of varying morphology at z=0, finding parallel sequences for the different Hubble types. We restrict our comparison sample to objects whose bulge to disk ratio is smaller than 0.3, which according to the authors ensure the selection of late-type galaxies (Sa to Sd). After applying this selection criterion, this sub-sample is composed of 44 disc galaxies within the range 9.0<$\log M_*$<11.2 for individual objects. Measurements in the local universe established a zero point to any scaling relation evolutionary path across cosmic time. To ensure that we can reliably fix this zero point, we included a second local universe comparison sample by \cite{Posti18}, who revisited previous $z=0$ studies on angular momentum by using a sub-sample of 92 nearby galaxies from the SPARC survey (\citealt{Lelli16}). We follow the same late-type selection criteria we applied to the \cite{Fall13, Fall18} sample and remove 16 galaxies with S0, irregular or compact morphology to end up with a sub-sample of 76 spiral galaxies with stellar-mass values ranging 8.0<$\log M_*$<11.2. 

Our third field comparison sample is composed of galaxies from the KROSS survey (\citealt{Stott16}) at z$\sim$0.9. Our selection criteria follow the approach of \cite{Harrison17} in their angular momentum study whilst adding tighter constraints. We decided to only use galaxies that are rotationally supported (i.e. V$_{rot}/\sigma$>1) and with well-determined effective radius and inclination angles from imaging data (i.e. quality 1 in \cite{Harrison17}). In addition, we discard those galaxies whose inclination angles are lower than 25º due to the high systematic errors that small variations in this parameter may introduce in the determination of the rotation velocity. After this, the KROSS sample is made of 301 objects within a mass range of 8.7<$\log M_*$<10.0. 

At high redshift, we use the KMOS Galaxy Evolution Survey (KGES, \citealt{Gillman20}) and the SINS/zC-SINF survey (\citealt{Forster-Schreiber18}) to explore the angular momentum evolution of field galaxies. The first sample is composed of 25 main sequence star-forming disc galaxies at $z\sim1.5$ published in \citealt{Gillman20}. These galaxies display a mass range given by 9.5<$\log M_*$<11.1. The SINS sample was originally composed of 35 galaxies from which we remove 7 objects because of their irregular morphological classification and their insufficient rotational support (i.e. V$_{rot}/\sigma$<1). We discard 3 additional objects due to their relatively lower redshift (z$\sim$1.5) in comparison with the rest of the sample. Thus, we end up with a sample of 25 disc star-forming galaxies at $2<z<2.5$ with stellar masses in the range of 9.3<$\log M_*$<10.5.

While the stellar-mass range of our cluster samples at intermediate to high redshift is cut-off at $\log M_*=9.5$, the field galaxy samples previously introduced display a significant population of galaxies below this threshold, even down to $\log M_*=8.0$ for the local studies. The absence of this kind of galaxies in our cluster samples can be explained by the moderate size of our cluster observing programs compared to some the field studies, and by the difficulty to detect and extract kinematic information from low-luminosity, low-mass objects, especially in the outer parts of a rotation curve. To tackle this issue, we restrict our subsequent comparative analysis between the cluster and field samples to galaxies above $\log M_*=9.5$. After applying this mass cut, the distribution of our field comparison samples with respect to their $M_*$, $V_{max}$, and $R_e$ can be found in Fig. \ref{F:Hist_field}, while their mean values are summarized in Table \ref{T:Hist_samples}. We note that the mean and median values for these three parameters agree within their typical errors (i.e. 0.1 dex for $\log (M_*)$, 30 km/s for $V_{max}$, and 0.3 kpc for $R_e$) for both our cluster and field comparison samples.
  
\section{Parameters of CL1604 galaxies} 
\label{SS:Methods}
  
\subsection{Rest frame magnitudes and stellar masses}
\label{SS:Imaging}

We use the SED fitting code {\sc Lephare} (\citealt{Ilbert2006}, \citealt{Arnouts2011}) to compute the rest-frame magnitudes and stellar masses in our cluster sample. For every object, {\sc Lephare} fits the spectral energy distribution given by the available photometric bands to a library of stellar population synthesis models (\citealt{BC03}) assuming Calzetti's attenuation law (\citealt{Calzetti2000}). We constrained the models to use extinction values of $E(B-V)=0 - 0.5$\,mag in steps of 0.1\,mag, and to produce galaxy ages lower than the age of the Universe at z$\sim$0.9 (i.e. 6.2 Gyrs) assuming a Chabrier IMF (\citealt{Chabrier03}). Our rest-frame magnitudes and logarithmic stellar masses are computed this way to an accuracy of 0.1 magnitudes and 0.15 dex respectively. To study the redshift evolution of the Tully-Fisher relation in Sect.\,\ref{SS:Results}, we must correct the derived absolute B-band magnitudes for extinction due to their inclination angles with respect to the line of sight. In edge-on spirals, the stellar light has to travel through the galaxy disc, that is filled with dust particles, before reaching us. Therefore, these galaxies possess higher extinction values than their face-on counterparts. Besides, more massive galaxies have higher dust content than low-mass objects \citep{Giovanelli95}, which introduces a stellar-mass dependence in the inclination extinction correction. We take into account these two effects following the prescription given by \citet{Tully98}. This correction diverges for completely edge-on galaxies (i.e. $i\approx90º$), and therefore we exclude them from our sample. After applying this correction, the typical errors for the B-band absolute magnitude values for the CL1604 sample are $\sim$0.2 mag.

\subsection{Structural parameters}
\label{SS:Structural}  
Space-based HST observations are ideal to measure the structural parameters of our targets reliably due to their high spatial resolution and depth. The field where CL1604 resides is covered by extensive HST imaging in two filters (F606W and F814W) covering most of the structure of the cluster complex, including all but one of our targets. In general, images taken in redder filters capture the light from the old stellar population that dominates the structure of the galaxy, which diminishes the contamination coming from prominent star-formation regions. We chose the F814W images as the main source to measure the structural parameters of our cluster members for this reason. We also use the HSC z-band to make the same measurements over the single object that has no HST imaging due to the depth and seeing conditions (FWHM$\sim$0.5") achieved in this band during the observations.

We model the surface brightness profile of our targets and measure their structural parameters by using the GALFIT code (\citealt{Peng02}). The models are produced following a two-component approach: First, we fit a single component exponential profile (Sèrsic index $n_{s}=1$) to every galaxy and subtract it from the original image. After visually inspecting the residuals we determine if the object under scrutiny shows signs of a bulge presence. If this is not the case, we keep the structural parameters computed with a single exponential disc surface brightness profile. However, if there are strong residuals in the central areas of the galaxy after the model subtraction, we proceed with a simultaneous 2-component fitting by adding another surface brightness profile with $n=4$ to take into account the bulge contribution while keeping the single component results as the initial guess values for the first component.

The modeling provides us with the position angle of our objects ($\theta$) with respect to the north direction, the effective radius ($R_{e}$) of the disc component, and the ratio between the apparent minor and major axis ($b/a$). The position angle can be used to identify possible misalignments between the major axis of the galaxy and its slit. This quantity is called the mismatch angle ($\delta$) and will be used at a later stage to correct the observed rotation velocities of our targets. The ratio between the axes, $b/a$, can be used to compute the inclination ($i$) of the galaxy with respect to the line of sight, which also plays an important role in the determination of the maximum rotation velocity. Finally, spiral galaxies have a small although significant scale height ($q$) that enters in the determination of $i$ following the approach given by \cite{Heidmann72}:
\begin{equation}
    \cos{(i)}=\sqrt{\frac{(b/a)^2-q^2}{1-q^2}}
\end{equation} 
 where $q=0.2$ represents the typical observed value for local spiral galaxies (\citealt{Tully98}). The three clusters that belong to our primary sample (RXJ1347, CL1604, and XMM2235) rely on $R_{e}$ measurements from different filters and at different redshifts. However, \cite{Jong96} found that the effective radius of disc galaxies experiences significant variation with wavelength. \cite{Kelvin12} measured a reduction of 25\% in $R_{e}$ for late-type galaxies from g to K-band, and established a relation accounting for these changes using measurements from the GAMA survey: 
\begin{equation}
    \log r_{e,disc} = -0.189 \log \lambda_{rest} +1.176 
\label{E2}
\end{equation}
where $\lambda_{rest}$ is the observed rest-frame wavelength for the galaxy. We use this relation to normalize the $R_{e}$ measurements of our three cluster galaxies to the same reference wavelength, $\lambda_{ref}\sim8100\AA$. This value is the rest-frame central wavelength of the I-band Johnson filter which was used to characterize the local velocity-size relation in \cite{Haynes99a} and by \cite{Boehm16} for their $z<1$ field sample of galaxies. After applying this correction the galaxies' effective radius decreases by $\sim10\%$ for the RXJ1347 sample (observed with SUBARU Suprime-Cam/z' at z$\sim$0.45), and $\sim19\%$ for the CL1604 sample (observed with HST/F814W at z$\sim$0.9), while it increases around $\sim5\%$ for the XMM2235 sample (observed with HAWKI/K-band at z$\sim$1.4). Our previous analysis of HST/ACS images of disc galaxies reported systematical errors of 20\% in galaxy sizes (\citealt{Boehm13}, \citealt{JM20}). We adopt this error value for our work in the following.

\subsection{Determination of the maximum intrinsic velocity ($V_{max}$)}
   
Our approach to extract the rotation curve from prominent emission lines and the subsequent modeling to determine the maximum rotation velocity ($V_{max}$) has been extensively described in previous publications from our group (see \citealt{Boehm04}, \citealt{Bosch2}, \citealt{Boehm16}). However, we provide a summary of the most important steps of the process in the following paragraphs for the readers' convenience.

First, we find our prominent spectral feature within the 2D spectra ([OII]3727$\AA$ for CL1604)) and measure the central wavelength position of the emission line by fitting a Gaussian profile over it row by row. We average the emission line over three neighboring rows (i.e. 0.75" in the spatial axis) to enhance the signal-to-noise (S/N) before the fitting. For every row, we inspect the small blue- and redshifts of the central wavelength position in the dispersion axis and transform them into positive and negative velocities with respect to the systemic velocity at the center of the galaxy. This way we obtain a position-velocity diagram that displays the rotation velocities as a function of galactocentric radius. We allow for small variations between the photometric and kinematic center of the galaxy of up to $\pm1$pix, i.e. $\sim$2kpc in spatial scale at z$\sim$0.9. 

The second step of the process involves the correction of the position-velocity diagram from all observational (beam-smearing) and geometrical effects (inclination, misalignment angle, and slit width) that may affect the observed values. To solve this, we generate synthetic velocity fields assuming an intrinsic rotation law, taking into account the seeing conditions at the time of the observations and the structural parameters previously determined through surface brightness modeling. We follow the multiparametric rotation law presented in \cite{Courteau97} which is characterized by a linear rise at distances smaller than the turnover radius ($r_t$) and a constant maximum rotation velocity ($V_{max}$) beyond this point, where the dark matter halo dominates the mass distribution. Finally, we place a slit along the major axis of the object and extract the synthetic rotation velocity values from the model as a function of radius. These values define a synthetic rotation curve that is allowed to change by tuning the $V_{max}$ and $r_t$ to fit (via $\chi^{2}$ minimization) the observational shape directly extracted from the 2D spectra. The precision achieved in the determination of $V_{max}$ is mainly influenced by the accuracy of the structural parameters (especially $i$) and the quality and extent of the rotation curve, with typical error values around $20$ km/s. In CL1604, only 8 regular rotation curves could be extracted out of 34 observed cluster objects, 12 of which displayed irregular kinematics. The remaining galaxies showed just gradients or too compact emission to asses their kinematic state. The synthetic and observed rotation curves can be found in Appendix\,\ref{A:app} for Cl1604 cluster members. The same information is available for XMM2235 and RXJ1347 in our past publications (\citealt{JM17, JM20}).

\subsection{Star-formation activity}  
\label{SS:sSFR}

Between the wide variety of star-formation diagnostics available from different emission lines, the H$\alpha$ prescription developed by \cite{Kennicutt92} has proven to be one of the most reliable (\citealt{Moustakas06}). However, in the intermediate-to-high redshift regime, the access to the H$\alpha$ emission-line can only be achieved through NIR observations. The three clusters that form part of our cluster primary sample (i.e. RXJ1347, CL1604, and XMM2235) only have spectroscopy in the optical wavelength range, which forces us to use a different calibration. The [OII] emission line doublet at $3727\AA$ has also been used as a proxy to estimate the SFR of galaxies empirically calibrating this indicator with respect to the H$\alpha$ diagnostic (e.g. \citealt{Verdugo08}). However, the [OII] calibration is subject to significant uncertainties related to the chemical evolution of galaxies, the dust reddening and the ionizing process of star-forming galaxies (\citealt{Moustakas06}, \citealt{Kennicutt12}). 
All these effects are of special importance when examining galaxies at high redshift as the available information in the rest-frame optical wavelength range only covers the bluer parts of the galaxies' spectra. Despite these caveats, we use the $SFR_{[OII]}$  diagnostic outlined in \cite{Kennicutt92} to roughly estimate the SFR of our sample: 

\begin{equation}
    SFR_{[OII]}/(M_{\odot}{yr}^{-1})=2.7\times10^{-12}\frac{L_B}{L_{B,\odot}}EW([OII])E(H_{\alpha})
\label{E22}
\end{equation}

where $(L_B/L_{B,\odot})=10^{0.4 (M_B-M_{B,\odot})}$, $M_{B,\odot}=5.48$ mag, EW([OII]) is the equivalent width of the [OII] emission line and $E(H_{\alpha})$ is the extinction around $H_{\alpha}$. We estimate the value of this quantity by assuming a \cite{Calzetti2000} extinction law, a \cite{Chabrier03} IMF, and taking the $E(B-V)$ reddening value obtained from the stellar continuum SED fitting (see Sect. \ref{SS:Imaging}) which accounts for the diffuse dust attenuation in the galaxy ($A_{cont}$). However, this value should be corrected by adding an extra contribution to the attenuation coming from the active star-forming regions ($A_{extra}$). We follow the approach outlined in \cite{Wuyts13} who studied the dust attenuation of galaxies at $0.7<z<1.5$ finding that $A_{extra}=0.9A_{cont}-0.15A_{cont}^{2}$, in good agreement with the previous estimates made by \cite{Calzetti2000} in the local universe.

\begin{figure}
\centering
\includegraphics[width=\linewidth]{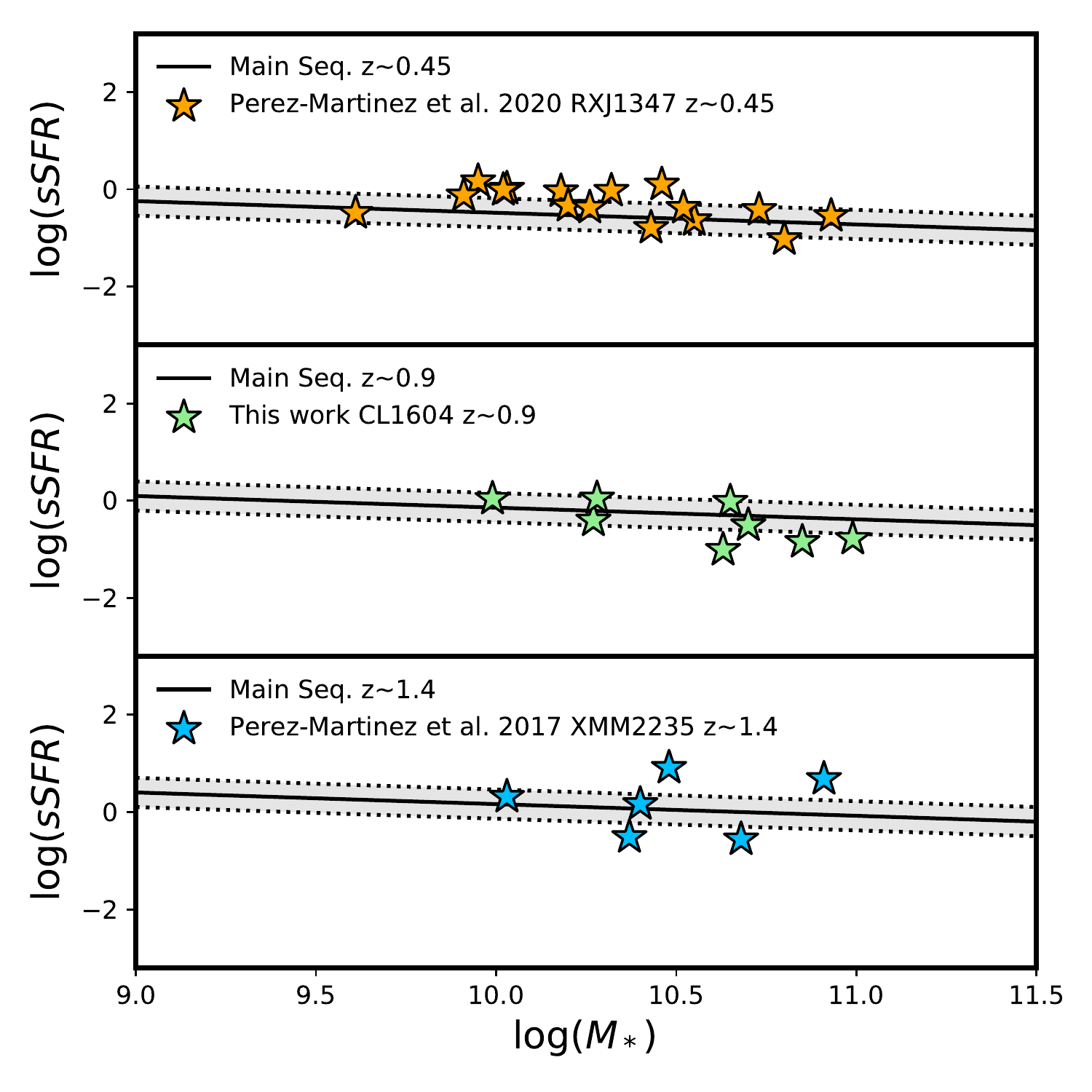}\par 
\caption{Star-formation activity as a function of stellar mass. Orange, green and blue stars respectively display cluster galaxies within RXJ1347, CL1604 and XMM2235. The solid line shows the main sequence of star-forming galaxies at $z\sim0.45$, $z\sim0.90$, $z\sim1.4$ extrapolated by \cite{Peng10} with a gray area indicating the 3$\sigma$ scatter.}
\label{F:SFR}
\end{figure}

The SFR and $E(B-V)_{SED}$ values of our primary cluster sample can be found in Tables \ref{T:CL16}, \ref{T:RXJ}, and \ref{T:XMM}. The [OII] emission line did not enter the observed wavelength range for three objects in the RXJ1347 sample. These objects have been excluded from all diagrams involving SFR. \cite{Kennicutt92} estimated that the systematic uncertainty of using the [OII] calibration ranges from 30\% for local and low redshift samples to up to a factor 2-3 (0.3-0.5 dex) at high redshift. We display the specific star formation (sSFR) of our cluster samples in Fig.\,\ref{F:SFR} using the main sequence (Eq. 1 in \citealt{Peng10}) at the redshift of our clusters as a reference for the expected sSFR value in the field with an scatter equivalent to 0.3 dex (grey area). Most of our objects show sSFR values compatible with those of the main sequence in the field within the errors.


\subsection{The environment} 
\label{SS:Environment}

In this work, we aim to investigate the effects of the environment on several kinematic scaling relations and their redshift evolution. To this aim, we have studied the main kinematic and structural parameters of several cluster samples across cosmic time in previous sections. However, cluster membership in all these samples is usually defined as an interval in redshift space around the value given for the whole cluster structure, which usually coincides with its BCG. Even though this may be sufficient to qualitatively disentangle the general field population of galaxies from objects residing in denser environments, we need a quantitative way to measure the environment in order to study its influence within these dense regions. 

\begin{figure}
\centering
\includegraphics[width=\linewidth]{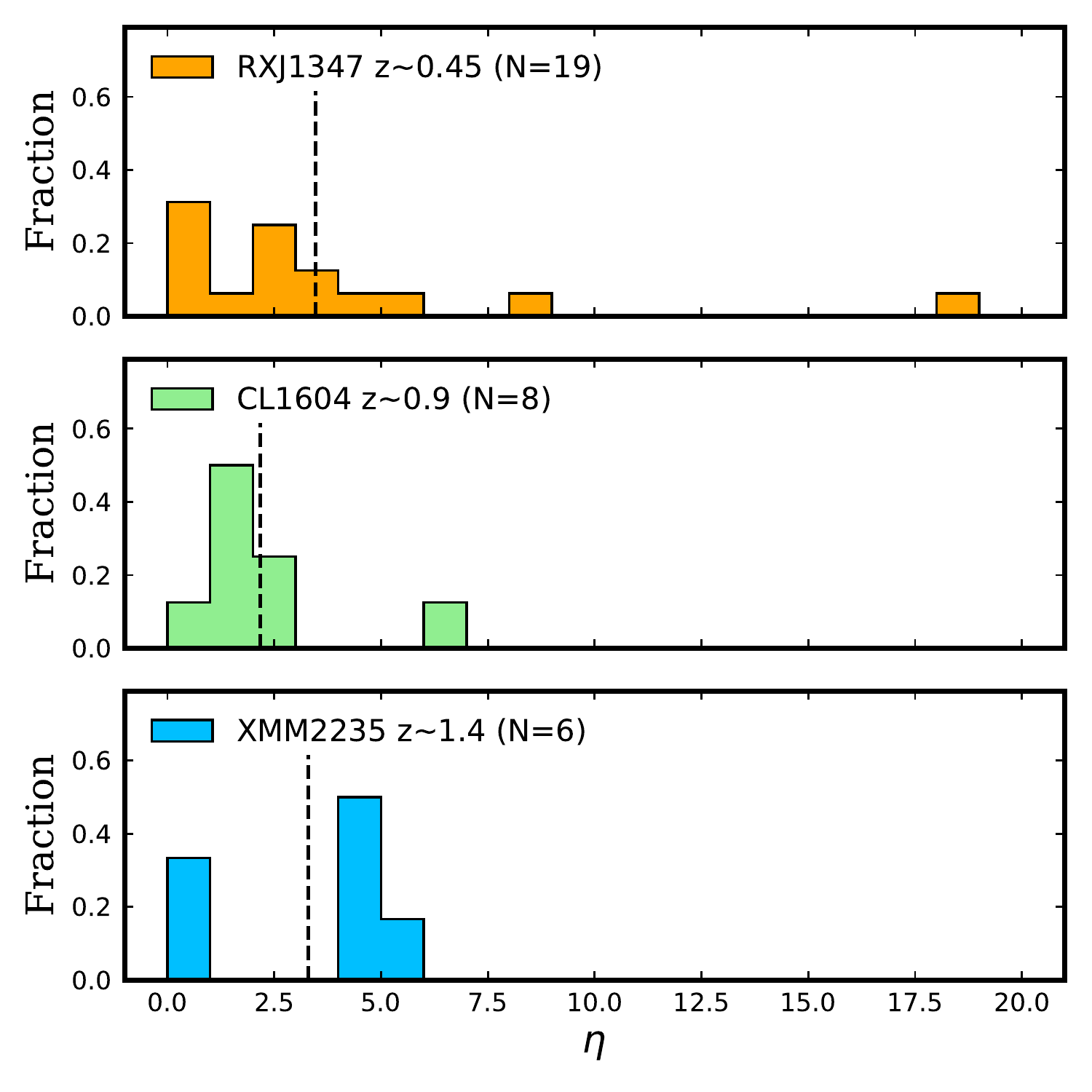}\par 
\caption{Global environment distribution of galaxies for RXJ1347, CL1604 and XMM2235 clusters. The $\eta$ value of each object has been computed according to its nearest cluster in Table \ref{T:clusters}. The dashed line displays the mean value of each sample.}
\label{F:eta}
\end{figure}

The environment can be quantified in two different ways: locally, as a number density of cluster members (e.g. \citealt{Dressler80}); and globally, by taking into account the general properties of the cluster ($M_{200}$, $R_{200}$, and $\sigma$) to define its phase-space (\citealt{Carlberg97}). While the first case requires a high number of known spectro-photometric cluster members to create a homogeneous mapping of each structure, the second one relies on the projected clustercentric distance ($R_{proj}$) of each object and its relative line-of-sight velocity with respect to the systemic velocity of the cluster ($\Delta v$), which can be measured through each object's redshift. We choose this second approach for our study given the heterogeneous origin of our samples, which prevents us to make a systematic study of the local environmental conditions within each cluster. We follow the approach outlined by \cite{Noble13} who used a parameter ($\eta$) that defines caustic profiles in a phase-space diagram in the following way:

\begin{equation}
    \eta=(R_{proj}/R_{200})\times(\left | \Delta v \right |/\sigma)
\label{E222}
\end{equation}

where $\left | \Delta v \right |= \left |(z-z_{cl})\ c/(1+z_{cl})\right |$ and $z_{cl}$ is the redshift of the cluster. Attending to this parameter, \cite{Noble13} defined three separate regions: $\eta<0.4$ for galaxies in the virialized region of the system, $0.4<\eta<2$ for galaxies that have been recently accreted, and $\eta>2$ for galaxies not yet associated with the main structure of the cluster. However, this scheme only traces the environment within virialized clusters while other minor cluster-related sub-structures such as filaments or infalling groups can not be accounted for. Thus, we decided to use $\eta$ as a continuous parameter that models the environmental relation of a given galaxy to a large cluster (according to Table \ref{T:clusters}) while keeping the objects with $\eta>2$, as they may form part of some minor sub-structures in the outskirts of the clusters. We display the environmental ($\eta$) distribution of our primary sample in Fig. \ref{F:eta} while the individual values are shown in Tables \ref{T:CL16}, \ref{T:RXJ}, and \ref{T:XMM} for each cluster. The mean $\eta$ values of each sample are equal to $3.5\pm1.0$, $2.2\pm0.7$, and $3.3\pm1.0$ for RXJ1347, CL1604 and XMM2235 respectively.

\section{Results}
\label{SS:Results}

The flat rotation curves of spiral galaxies provide us with a proxy, the rotation velocity ($V_{rot}$), to trace the total mass of the galaxy (including dark matter) as well as to study its relation with several other baryonic parameters. Some of the key additional parameters that describe the physics of spiral galaxies are the disk size (through the effective radius, $R_e$, or scale length, $R_d$), and the stellar population content of the galaxy (via its luminosity or stellar-mass). The rotation velocity, the galaxy size and the luminosity (or stellar mass) comprise a three-dimensional space (\citealt{Koda00}) that can be projected onto several planes to produce different scaling relations such as the Tully-Fisher relation (TFR) and the velocity-size relation (VSR, \citealt{Tully77} in both cases). Furthermore, different combinations between these parameters provide us with other interesting relations such as the specific angular momentum-stellar mass (AMR) that are key to understand the processes of morphological transformation and mass redistribution that galaxies suffer during their lifetime. In this work we use the cluster and field samples introduced in Sect.\,\ref{S:Sample} to study the evolution of the B-band Tully-Fisher, the Velocity-Size, and the angular momentum relation with respect to environment and time. In the first two cases, we will focus on samples exclusively studied by our group to achieve full consistency in the methodology to extract the main physical parameters between data-sets. For the angular momentum, on the other hand, we choose additional comparison field samples from the literature that provide the required parameters for their study (i.e. $M_*$, $R_e$ and $V_{max}$).

\subsection{The B-band Tully Fisher relation}

\begin{figure*}
    \begin{multicols}{2}
      \includegraphics[width=\linewidth]{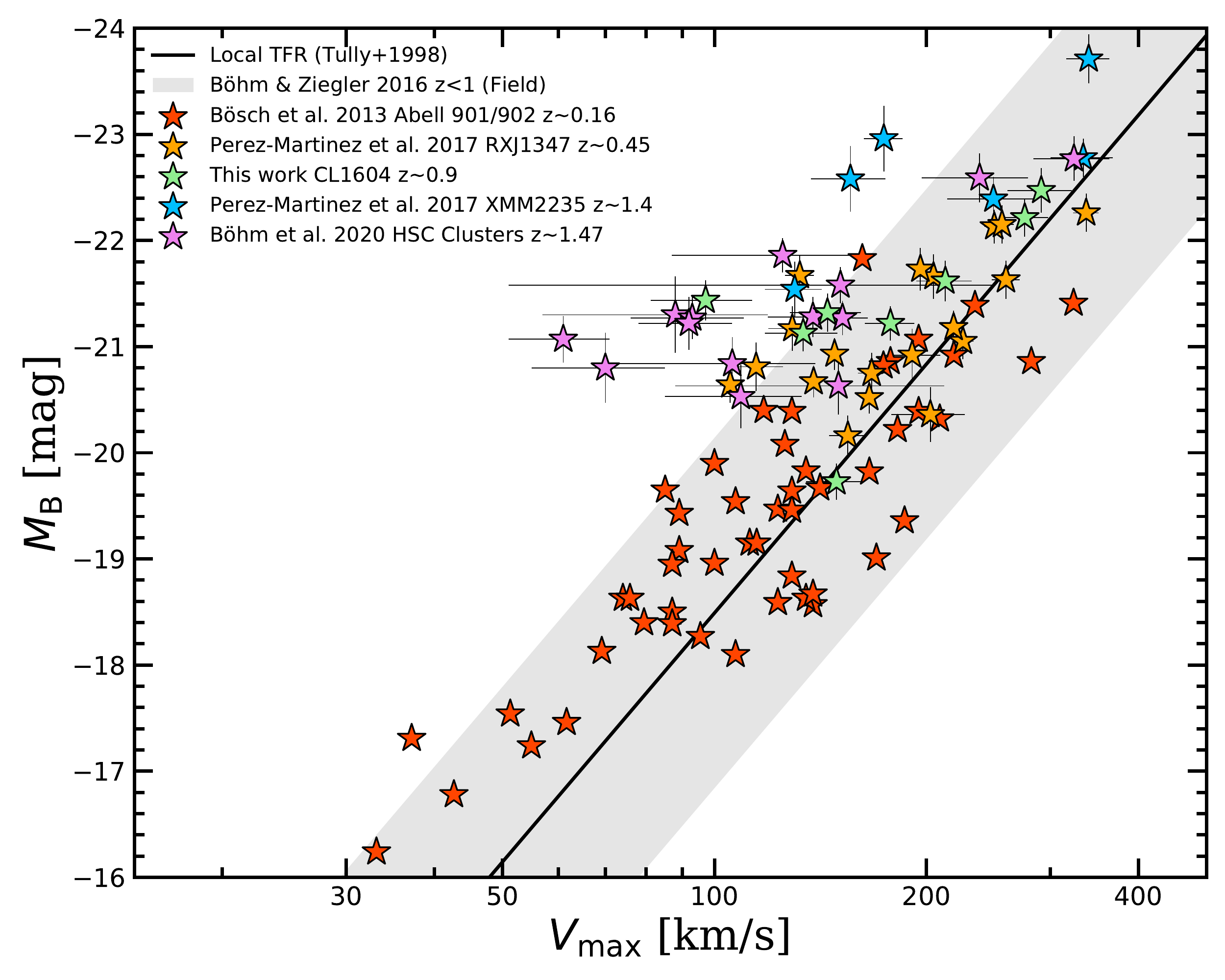}\par 
      \includegraphics[width=\linewidth]{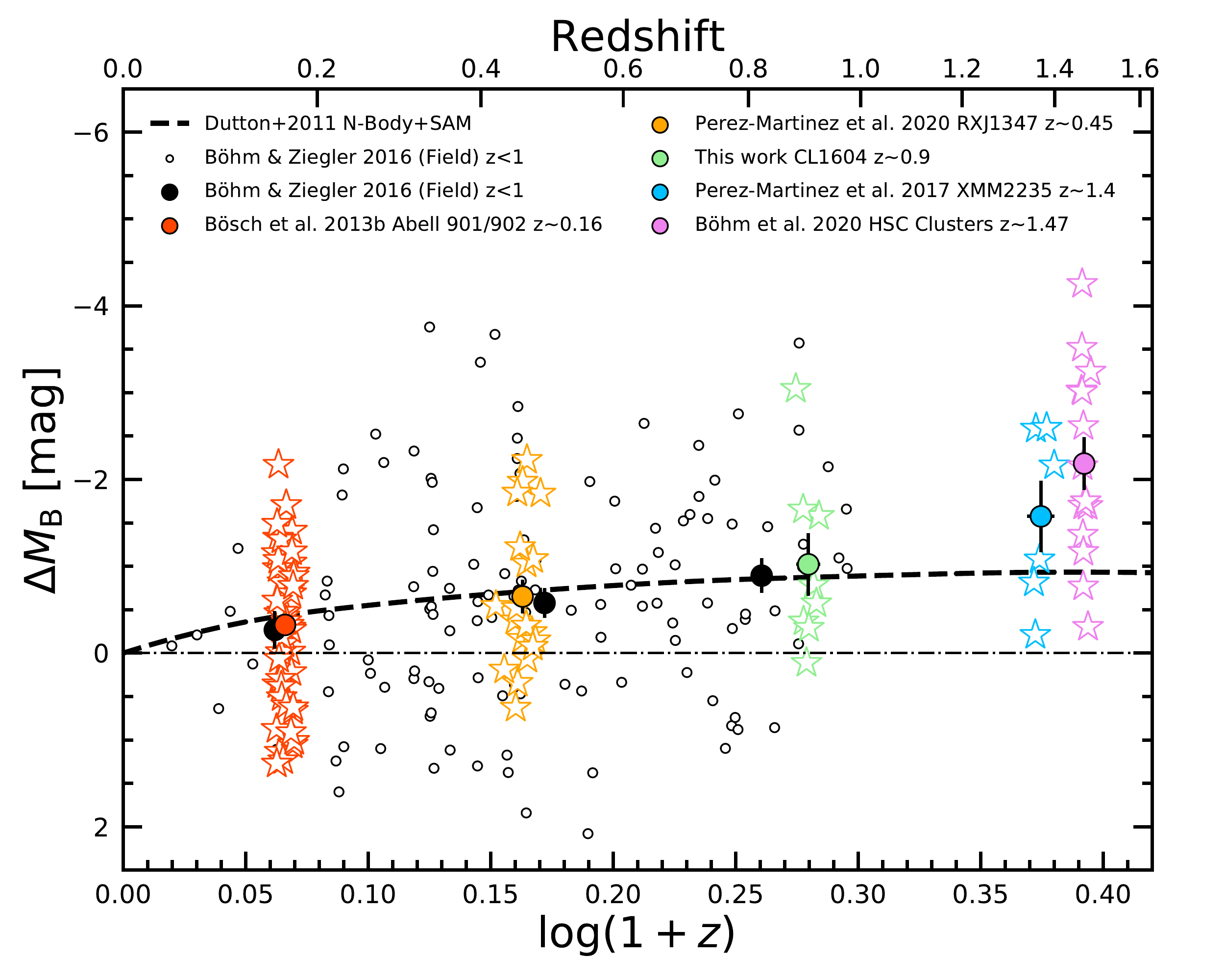}\par 
      \end{multicols}
      \caption{\textit{Left}: B-band Tully-Fisher relation. Colored stars represent the different cluster samples that compose our study. The black solid line shows the local B-band TFR (\citealt{Tully98}) with a 3$\sigma$ scatter area around reported by \cite{Boehm16} for galaxies at $0<z<1$ (grey area). \textit{Right}: B-band Tully-Fisher offsets evolution. Stars with colored edges represent the distribution of our cluster samples with their respective mean values being shown as bigger circles of the same color. Error bars represent the error of the mean for every sample. Open} circles display the \cite{Boehm16} field sample at $0.1<z<1$. We binned the field sample in three redshift intervals ($0<z<0.33$, $0.33<z<0.66$, $0.66<z<1$). Black circles depict the mean value and its error for the field sample in every redshift window. The dashed line represents the predicted B-band luminosity evolution by \cite{Dutton11} in the TFR, while the dashed-dotted line at $\Delta M_B=0$ means no size evolution.
         \label{F:TFR}
      \end{figure*}

First, we examine the distribution of our targets in the B-band Tully-Fisher diagram (Fig.\,\ref{F:TFR}, left-hand panel). Cluster objects are plotted using stars following the color scheme of previous figures to express their membership to different clusters. We use the local TFR (solid black line, \citealt{Tully98}) and a sample of 124 field disc-like galaxies from \cite{Boehm16} at $0.1<z<1$ for comparison. The grey area in Fig.\,\ref{F:TFR} depicts the distribution of the field sample around the local relation. Its half-width is equal to three times the scatter of the distribution (3$\sigma_{f}$), which encompass the majority of the field galaxies. 

Most cluster galaxies lie within the 3$\sigma_{f}$ field distribution, although at a fixed rotational velocity, the objects progressively move towards higher B-band absolute magnitudes with redshift. This is a natural consequence of the gradual evolution of the stellar populations with lookback time, with higher SFRs and younger (and hotter) stars contributing more to the luminosity of the galaxy when the universe was at an earlier stage. This effect can also be seen in the right-hand panel of Fig.\,\ref{F:TFR}, where we display the B-band magnitudes offsets from the local TFR ($\Delta M_B=M_{B,z}-M_{B,z=0}$) as a function of redshift. Due to the large scatter of our samples we also plot the mean offset values of our cluster datasets and their errors. We follow a similar approach for the field sample by splitting it into three redshift bins ($z<0.33$, $0.33<z<0.66$ and $0.66<z<1.0$). The mean B-band TFR offset value ($\overline{\Delta M_B}$), the error of the mean ($\sigma_{\overline{\Delta M_B}}$) and the dispersion ($\sigma_{\Delta M_B}$) for every sample can be found in Table \ref{T:Summary_1}.

 \begin{figure*}
    \begin{multicols}{2}
      \includegraphics[width=\linewidth]{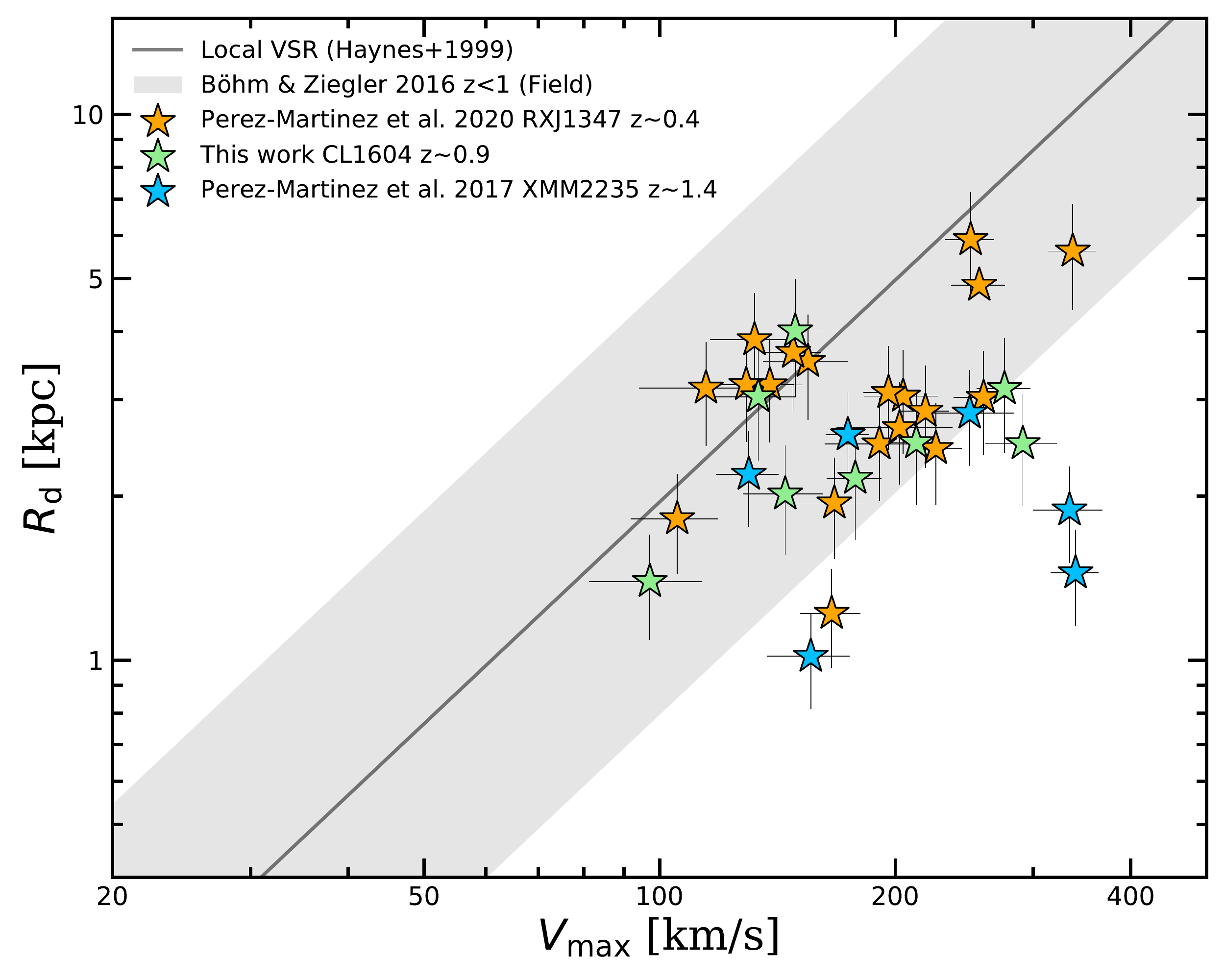}\par
      \includegraphics[width=\linewidth]{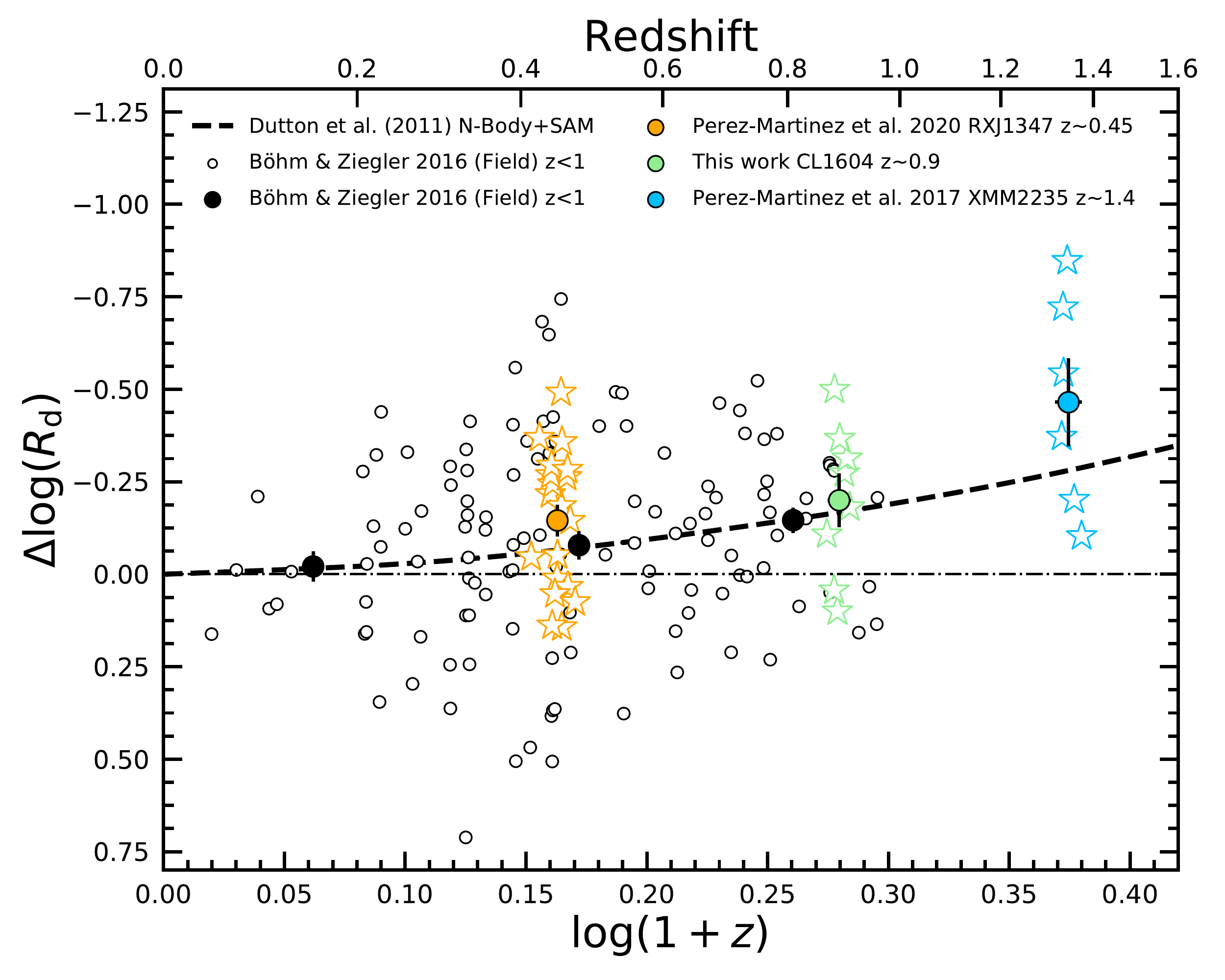}\par 
      \end{multicols}
      \caption{The symbols and color schemes are the same as in Fig. \ref{F:TFR}.} \textit{Left}: Velocity-Size diagram. The solid black line shows the local VSR from \cite{Haynes99b}, with a 3$\sigma$ scatter gray area around it. \textit{Right}: Velocity-size offset evolution.
         \label{F:TFR2}
      \end{figure*}
      
Our results show that there is a gradual increase in $\Delta M_B$ for cluster galaxies up to z$\sim$1 (colored circles, $\Delta M_B\approx1$). This trend is replicated by the binned mean values for field galaxies from \cite{Boehm16} (empty circles), which account for the same redshift intervals albeit with larger number statistics. Although the field mean values tend to lie slightly below the clusters' ones, this small difference becomes negligible when taking into account the errors of the mean. On the other hand, the semianalytical model by \cite{Dutton11} (dashed line) predicts a rise in B-band luminosity that is compatible with our results in the cluster as well as in the field at a similar redshift. This model is based on evolving dark matter haloes that host baryonic discs in their centers. The evolution of the stellar-mass and gas content of the discs is driven by the radial variation of star-formation, gas recycling, and accretion, which are the main processes that influence the evolution of galaxies in the field (\citealt{Peng10}). Thus, our results point towards little to no influence of the environment over cluster galaxies with regard to the B-band TFR up to z$\sim$1. However, the two higher redshift cluster samples significantly deviate from the semianalytical predictions at the 1.6$\sigma$ level for XMM2235 ($\Delta M_B=1.57\pm0.41$) and at the 4.2$\sigma$ level for the two HSC clusters ($\Delta M_B=2.18\pm0.30$). Unfortunately, we do not have a field comparison sample analyzed following the methods described in Sect.\,\ref{SS:Methods} at this high redshift. Nevertheless, this behavior hints to the presence of unaccounted processes influencing the B-band luminosity of high redshift cluster galaxies.

\subsection{The Velocity Size relation}
      
Galaxies display lower sizes at higher redshifts (\citealt{Bouwens04}). This is a consequence of the growth of disks across cosmic time and it is one of the predictions of the hierarchical growth of structures (\citealt{Mao98}). By construction, our cluster samples are exclusively made of disk galaxies (see Sect.\,\ref{SS:Methods}). Therefore, the disc scale length ($R_d$) can be used to investigate the size evolution of our cluster samples in the VSR. Our results are shown in the left-hand diagram of Fig.\,\ref{F:TFR2}, which keeps the same symbol scheme used in the TFR, though marking now the local velocity-size relation from \cite{Haynes99b} with a solid line. Galaxies from Abell 901/902 at z$\sim$0.16 and from the two HSC clusters at z$\sim$1.5 are excluded due to the lack of size measurements in their respective parent studies (\citealt{Bosch1}, \citealt{Boehm20}). In general, the disc size correlates with the wavelength at which it is observed, with bluer wavelengths yielding larger $R_d$ (\citealt{Kelvin12}, \citealt{Vanderwel14}). Even though $R_d$ was originally measured using different photometric bands for every sample, we use Eq.\,\ref{E2} to re-normalize all size measurements to the same rest-frame wavelength and compare our results with the field reference sample of \cite{Boehm16}. 

As we did in the TFR, we explore the redshift evolution of the velocity-size relation in the right-hand diagram of Fig.\,\ref{F:TFR2}, and compare our results once again with the semi-analytical models of \cite{Dutton11}. Interestingly, we find that cluster and field galaxies have a similar average size evolution, with a factor 1.6 drop in size ($\Delta\log{R_d}=-0.20\pm0.07$ for CL1604) by $z\approx1$. This result agrees with the predictions of \cite{Dutton11}, although individual objects display large scatter around the mean. This may be related to the different formation ages of galaxies and their distinct evolutionary paths. Observational studies in the field such as \cite{Vanderwel14} find that the size growth of disk galaxies with redshift is given by $R_e\propto(1+z)^{-0.75}$, which yields a factor 1.6 growth between $z=1$ and $z=0$ at a fixed stellar mass, confirming our previous results. However, this empirical relation only predicts a factor 2 growth at z=1.5, in agreement with \citealt{Dutton11}, but in contrast with our results in XMM2235 that show smaller sizes by almost a factor 3 ($\Delta\log{R_d}=-0.46\pm0.12$). The mean offsets with respect to the VSR ($\overline{\Delta\log r_d}$), the error of the mean ($\sigma_{\overline{\Delta\log r_d}}$) and the dispersion ($\sigma_{\Delta\log r_d}$) for every sample can be found in Table \ref{T:Summary_1}.

\subsection{The angular momentum}  

\begin{figure*}
\centering
\includegraphics[width=12cm]{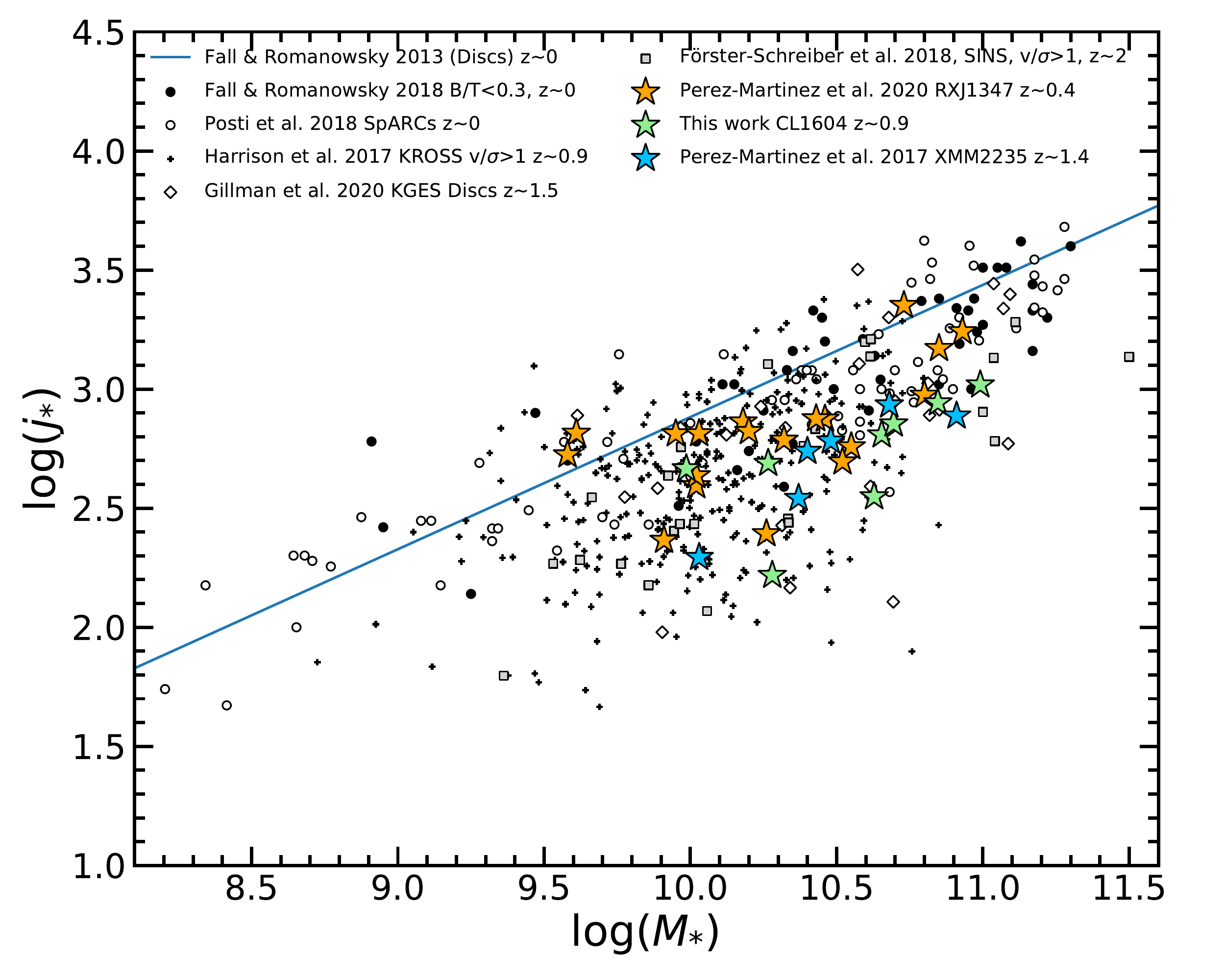}\par 
\caption{Specific angular momentum diagram. Orange, green and blue} stars respectively represent the RXJ1347, CL1604 and XMM2235 cluster galaxies studied by our group. Black and empty circles show the local universe disc-like samples from \cite{Fall18} and \cite{Posti18} respectively. The small black crosses show the field objects from KROSS sample at z$\sim$0.9 (\citealt{Harrison17}), empty diamonds display the KGES sample at z$\sim$ 1.5 (\citealt{Gillman20}), while the grey squares depict SINS/zC-SINF galaxies at z$\sim$2 (\citealt{Forster-Schreiber18}). The blue solid line display the local "Fall relation" for disc galaxies from \cite{Fall13}.
\label{F:AM}
\end{figure*}

The angular momentum ($J$) simultaneously connects all the relevant parameters involved in the previous scaling relations, i.e. stellar-mass, size, and rotation velocity. The transference of angular momentum from the dark matter halo to the baryonic component is key to understand the early stages of galaxy formation (\citealt{Mo98}). On the other hand, the specific angular momentum defined as j$_*$=J/M$_*$ has proven to be a fundamental quantity to explore galaxy evolution and morphological transformation. \cite{Fall83} first found a tight relation between $j_*$ and M$_*$ with the form $j_* \propto M^{2/3}$. Its normalization depends on the galaxy morphological type, with parallel sequences towards lower $j_*$ values for early-type galaxies, pointing towards a loss of angular momentum of up to an order of magnitude linked to the morphological evolution of galaxies (\citealt{Romanowsky12}, \citealt{Fall13,Fall18}). In this section we compare the results obtained for the specific angular momentum relation (i.e. j$_* - \log M_*$) for cluster and field galaxies at different epochs. To maintain consistency between the analysis of our cluster and field samples, we adopt the theoretical frame described in \cite{Harrison17} to study the evolution of angular momentum, for which we present a summary in the following paragraphs. First, equation 6 in \cite{Romanowsky12} will be used as an estimate for specific angular momentum:
\begin{equation}
    j_* = k_n C_i v_s R_e
\label{E3}
\end{equation} 
where $R_e$ is the effective radius of the galaxy, $v_s$ is the observed rotation velocity at some arbitrary radius, $C_i$ is an inclination correction factor and $k_n$ is a numerical factor that takes into account the current morphology of the galaxy approximated by its Sérsic index ($n$) in the following way:
\begin{equation}
    k_n = 1.15+0.029n+0.062n^2
\end{equation} 
By construction (see Sect.\,\ref{SS:Structural}), our samples only contain disk galaxies so that they display characteristic exponential surface brightness profiles ($n\sim1$). Furthermore, small variations of n in the vicinity of the exponential profile (for example $o.5<n<1.5$) will only introduce small variations in the value of $k_n$ (up to $7\%$). Therefore, we confidently assume $n=1$ as our standard value for all further calculations, adding such uncertainty contribution to the specific angular momentum error budget. Additionally, we may consider that our inclination and seeing corrected maximum rotation velocity is equivalent to $V_{max} \approx C_i v_s$, simplifying Eq.\,\ref{E3} for disc galaxies to just:
\begin{equation}
    j_* \approx 2V_{max}R_d
\end{equation} 
where $R_d$ is the disc scale-length and $R_d\approx R_e/1.678$. This approach converts the AMR in a correlation between two independent variables. In Fig.\,\ref{F:AM}, we present the results obtained for our cluster galaxies in comparison with our selection of field samples, which have been analyzed following the same method described above. The distribution of the local data is fitted by the expectations for disks given by \cite{Fall13} (blue line). At higher redshifts, both the field and the cluster samples display lower specific angular momentum values but seem to follow sequences with similar slope. However, the scatter of our data and the limited number statistics of the cluster members prevent us from computing a slope value that could be used as a comparison between the cluster and field samples at different redshifts.

To further explore the possible variations in specific angular momentum as a function of redshift and environment, \cite{Harrison17} applied a simple predictive model based on the works of \cite{Romanowsky12} and \cite{Obreschkow14} (their eq. 18 and 19), assuming that the baryonic fraction can be approximated by $f_b=0.17$ (\citealt{Komatsu11}):
\begin{equation}
    \frac{j_{*,pre}}{kpc\,km\,s^{-1}}=2.95 \cdot 10^4 f_j f_s^{-2/3} \lambda \left(\frac{H[z]}{H_0}\right)^{-1/3} \left[\frac{M_*}{10^{11}M_{\odot}}\right]^{2/3} 
\label{EQ6}
\end{equation}
where $H[z]=H_0(\Omega_{\Lambda}+\Omega_m [1+z]^3)^{1/2}$. This model relies on the assumptions that all galaxies reside inside singular isothermal spherical cold dark matter haloes characterized by a spin parameter $\lambda$ and a specific angular momentum $j_{h}$. Therefore, the galaxies embedded in such dark matter haloes possess a fraction of the specific angular momentum of the halo in which they were formed $f_j=j_*/j_{halo}$. This fraction may change between galaxies from different epochs and evolutionary paths since it represents the specific angular momentum retained by the baryons at a given moment, and will be the focus of our analysis. 

\begin{figure}
   \centering
   \includegraphics[width=\linewidth]{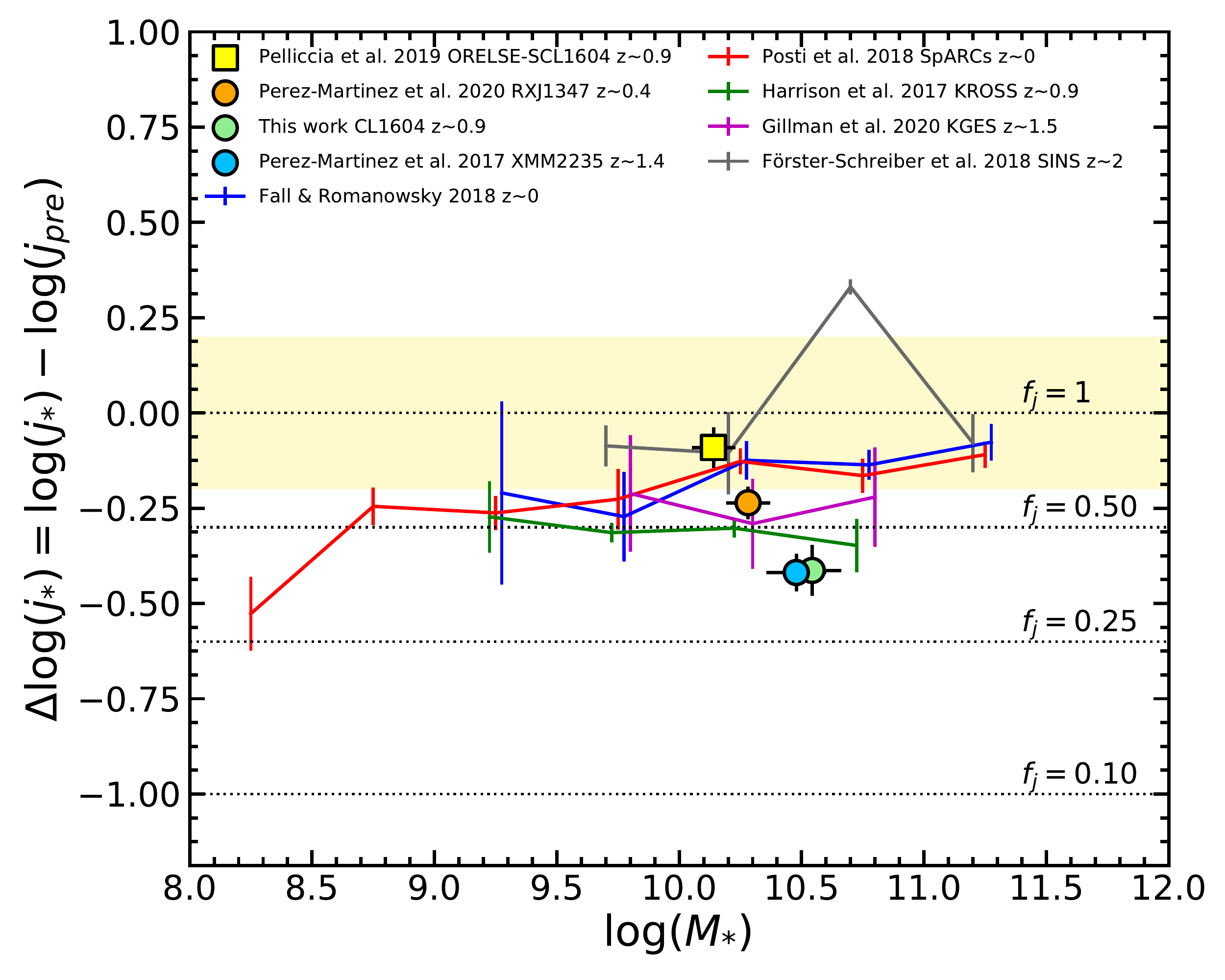}\par
   \caption{Specific angular momentum offsets ($\Delta\log(j_*)$) as a function of stellar mass. The solid colored lines depict the mean values and the error of the mean of the mass-binned field samples using bin widths of 0.5 dex. Blue and red lines depict the local universe disc samples from \cite{Fall18} and \cite{Posti18} respectively. The green line shows the distribution of objects from the KROSS sample at z$\sim$0.9 (\citealt{Harrison17}). The purple line shows the z$\sim$1.5 KGES sample of galaxies (\citealt{Gillman20}), while SINS/zC-SINF galaxies at z$\sim$2 are displayed by the gray line (\citealt{Forster-Schreiber18}). The dotted lines mark different predicted values of $f_j$ as a function of $\Delta\log(j)$. The light yellow band depicts the predicted scatter in the spin parameter $\lambda$ assuming $f_j=1$. The orange, green and blue circles respectively show the mean $\Delta\log(j)-\log(M_*)$ values and their errors for RXJ1347, CL1604 and XMM2235 cluster galaxies. Finally, the yellow square display the average value for \cite{Pelliccia19} cluster sample.}
   \label{F:AM_OFFSET}
\end{figure}

During the processes of galaxy formation, the asymmetric collapse of high-density regions generates tidal torques that introduce a particular angular momentum value for every mass distribution (\citealt{Hoyle51}, \citealt{Peebles69}). The spin parameter accounts for this behavior. N-body simulations and recent observational studies have shown that $\lambda$ follows a near-lognormal distribution with an expected value $\lambda=0.035$ and a mean dispersion of 0.2 dex, which remains approximately constant when examining different epochs, galaxy masses, and environments (\citealt{Maccio07, Maccio08}, \citealt{Romanowsky12}, \citealt{Bryan13}, \citealt{Burkert16}). Finally, $f_s$ represents the stellar mass fraction relative to the initial gas mass. This parameter is a function of the different internal processes that take place within a galaxy as a result of its evolution. Given the difficulty to account for all possible variables involved in the baryonic physics, \cite{Harrison17} followed an empirical approach and use the mass-dependent description given by \cite{Dutton10} for late-type galaxies and revisited in \cite{Burkert16} for this purpose:
\begin{equation}
    f_s=0.29\left(\frac{M_*}{5\cdot 10^{10}M_{\odot}}\right)^{1/2}\left(1+\left[\frac{M_*}{5\cdot 10^{10}M_{\odot}}\right]\right)^{-1/2} 
\end{equation}
Under these assumptions, the only free parameter in $j_{*,pre}$ is the specific angular momentum fraction retained by the galaxy with respect to its dark matter halo ($f_j=j_s/j_{halo}$). We now can consider the idealized case where the baryonic and dark matter component of the galaxy have been well-mixed from the early stages of galaxy formation, meaning that the specific angular momentum in both components is very similar ($j_*\approx j_{halo}$), and thus, $f_j\sim1$. In Fig.\,\ref{F:AM_OFFSET} we present the discrepancies between the predicted specific angular momentum (assuming $f_j=1$) and the values computed following Eq.\,\ref{E3} for all the cluster and field galaxies (i.e $\Delta\log(j_*)=\log(j_*)-\log(j_{pre})$) as a function of stellar mass. This choice of parameters is very useful because the negative values in $\Delta\log(j_*)$ can be re-interpreted as lower fractions of conserved specific angular momentum (i.e. $f_j<1$) for galaxies that have experienced different conditions (e.g. redshift or environmental evolution). Thus, it provides a way to directly compare the amount of conserved angular momentum ($f_j$) as a function of stellar-mass for galaxies with diverse origins. 

In Fig.\,\ref{F:AM_OFFSET} we show the behavior of the field objects in $\Delta\log(j_*)$ by binning these samples within intervals of 0.5 dex in $M_*$. The solid lines show the resulting mean values and their errors as a function of stellar mass. Objects above $\Delta\log(j_*)=0$ (i.e. $f_j>1$) are allowed to exist in this model due to the uncertainty in the determination of the spin parameter ($\lambda$) and its mean dispersion value (0.2 dex, yellow band). We avoid the binning of our cluster samples due to their poor sampling across the given mass range. Instead, we simply compute the mean values ($\overline{\Delta\log j_*}$), errors of the mean ($\sigma_{\overline{\Delta\log j_*}}$) and dispersion ($\sigma_{\Delta\log j_*}$) per cluster (colored circles), which can also be found in Table \ref{T:Summary_2}.

We obtain $\overline{\Delta\log(j_*)}=-0.24\pm0.04$ for galaxies in RXJ1347 at z$\sim$0.45, $\overline{\Delta\log(j_*)}=-0.41\pm0.07$ for galaxies in CL1604 at z$\sim$0.91, and $\overline{\Delta\log(j_*)}=-0.42\pm0.05$ for galaxies in XMM2235 at z$\sim$1.39. These three values indicate a significant increase in the angular momentum since z$\sim$1 but little evolution between z$\sim$1 and z$\sim$1.4. As we stated above, we can re-interpret these offsets as a decrease in the retained specific angular momentum fraction, $f_j$, with increasing redshift. Thus, galaxies from RXJ1347 on average conserve $58\pm5\%$ of their halo specific angular momentum (i.e. $\overline{f_j}=0.58\pm0.05$) while the cluster members from CL1604 and XMM2235 on average display $\overline{f_j}=0.39\pm0.07$ and $\overline{f_j}=0.38\pm0.05$ respectively. By comparison, we also include the cluster sample recently studied by \cite{Pelliccia19} within the CL1604 multicluster system. Out of the 94 galaxies that are part of their ORELSE-SC1604 sample, only 22 of them display $\log(M_*)>9.5$, $v/\sigma>1$, and can be considered to be part of the CL1604 cluster system according to the redshift limits imposed by \cite{Lemaux12} for this structure (0.84<z<0.96). We measure $\overline{\Delta\log(j_*)}=-0.09\pm0.05$ which corresponds with $\overline{f_j}=0.81\pm0.08$ for their sample. These values are compatible with those of the local samples at similar stellar mass ($\log(M_*)=10.0-10.5$ mass bin), where $\overline{\Delta\log(j_*)}=-0.12\pm0.05$ for the \cite{Fall18} sample and $\overline{\Delta\log(j_*)}=-0.13\pm0.03$ for the \cite{Posti18} sample. In contrast, the results of the KROSS sample at the same mass bin yields $\overline{\Delta\log(j_*)}=-0.30\pm0.02$ which implies a 4.2$\sigma$ difference between \cite{Pelliccia19} and the KROSS samples. This gap becomes even larger when compared with our high redshift cluster samples (Cl1604 and XMM2235), although they also display slightly higher mean stellar-mass values. This disagreement is maintained even when excluding galaxies with $\eta>3$ from the 22 selected galaxies within the ORELSE-SC1604 sample. However, \cite{Pelliccia19} also claim that methodological differences in their kinematic measurements may not allow to directly compare their results with others such as the KROSS survey.

We now compare our results in dense environments at different epochs with the field comparison samples. In the field at z$\sim$0 there is little dependence of $\Delta\log(j_*)$ on the galaxy's stellar-mass for $\log(M_*)>9.5$. The variations are negligible when taking into account the errors of the mean for every bin in the local samples (\citealt{Fall18} in blue and \citealt{Posti18} in red), which display similar average offset values across the mass range under scrutiny. 
Examining the selected KROSS and KGES disc galaxies at z$\sim$0.9 and z$\sim$1.5 (green and purple bins respectively) we find a similar situation with no stellar-mass dependence but an even lower mean offset across the chosen mass range. 
Similar results were previously reported in a larger sub-sample of the KROSS survey by \cite{Harrison17}, who applied less restrictive selection criteria. However, the errors of the KGES sample make it statistically consistent with the field samples in the local universe as well as at $z\approx1$. On the other hand, the $z>2$ galaxies from the SINS/zC-SINF survey (\citealt{Forster-Schreiber18}) display unusual small specific angular momentum offsets ($\overline{\Delta\log(j_*)}\approx-0.1$) compared to the other field z=1-1.5 samples (KROSS and KGES) but consistent with the z=0 field. This result is at odds with the expectation of specific angular momentum growth between z=2 and z=0 (see \citealt{Renzini20}, and references therein). The uncertainties in the measured sizes and morphologies (Sèrsic index values) reported in their study may be responsible for this disagreement as the method we used to estimate the specific angular momentum is only valid for disc (n$\approx$1) galaxies. In addition, possible differences between our methods to extract rotation velocities and those applied to the SINS sample may also contribute to the observed discrepancy.

\begin{table*}
\centering
\caption{Summary of the TFR and VSR properties of the cluster and field samples studied in this work. The columns represent: Sample IDs, redshift, mean B-band TFR offset, standard error of the mean B-band offset, standard deviation of the B-band offset values,  mean size offset in the VSR, standard error of the mean size offset and standard deviation of the size offset values.}

\begin{tabular}{cccccccc}
\hline
\noalign{\vskip 0.1cm}
Sample ID & z & $\overline{\Delta M_B}$ & $\sigma_{\overline{\Delta M_B}}$ & $\sigma_{\Delta M_B}$ & $\overline{\Delta\log r_d}$ & $\sigma_{\overline{\Delta\log r_d}}$ & $\sigma_{\Delta\log r_d}$ \\
\noalign{\vskip 0.1cm}
\hline 
\hline 
\noalign{\vskip 0.2cm}
Abell 901/902   & 0.16 & -0.32 & 0.12 & 0.86 & - & - & - \\
RXJ1347         & 0.45 & -0.65 & 0.19 & 0.84 & -0.15 & 0.04 & 0.19 \\
CL1604          & 0.91 & -1.02 & 0.36 & 1.02 & -0.20 & 0.07 & 0.20 \\
XMM2235         & 1.39 & -1.57 & 0.41 & 1.01 & -0.46 & 0.12 & 0.29 \\ 
HSC-Clusters    & 1.47 & -2.18 & 0.30 & 1.13 & - & - & - \\  
\noalign{\vskip 0.1cm}
\hline 
\noalign{\vskip 0.1cm}
\cite{Boehm16}  & 0.17 & -0.27 & 0.22 & 1.14 & -0.02 & 0.04 & 0.22 \\
...             & 0.50 & -0.58 & 0.17 & 1.36 & -0.07 & 0.04 & 0.31 \\
...             & 0.83 & -0.89 & 0.20 & 1.15 & -0.15 & 0.03 & 0.20 \\

\noalign{\vskip 0.1cm}
\hline 
\label{T:Summary_1}
\end{tabular}
\end{table*}

\begin{table*}
\centering
\caption{Summary of the angular momentum properties of the cluster and field samples studied in this work. Columns: Sample IDs, redshift, mean $\Delta\log j_*$ followed by its standard error and  standard deviation values,  mean $\log (j_*/M_*^{2/3})$ followed by its standard error and standard deviation values.
}
\begin{tabular}{cccccccc}
\hline
\noalign{\vskip 0.1cm}
Sample ID & z & $\overline{\Delta\log j_*}$ & $\sigma_{\overline{\Delta\log j_*}}$ & $\sigma_{\Delta\log j_*}$ & $\overline{\log (j_*/M_*^{2/3})}$ & $\sigma_{\overline{\log (j_*/M_*^{2/3})}}$ & $\sigma_{\log (j_*/M_*^{2/3})}$ \\
\noalign{\vskip 0.1cm}
\hline 
\hline 
\noalign{\vskip 0.2cm}
RXJ1347         & 0.45 & -0.24 & 0.04 & 0.19 & -4.04 & 0.05 & 0.22 \\
CL1604          & 0.91 & -0.41 & 0.07 & 0.19 & -4.31 & 0.07 & 0.20 \\
XMM2235         & 1.39 & -0.42 & 0.05 & 0.12 & -4.29 & 0.04 & 0.10 \\ 
\noalign{\vskip 0.1cm}
\hline 
\noalign{\vskip 0.1cm}
\cite{Fall18}   & 0   & - & - & - & -3.96 & 0.03 & 0.18 \\
SpARCs          & 0   & - & - & - & -3.97 & 0.03 & 0.21 \\
KROSS           & 0.9 & - & - & - & -4.10 & 0.02 & 0.30 \\
KGES            & 1.5 & - & - & - & -4.17 & 0.08 & 0.38 \\
SINS            & 2.0 & - & - & - & -3.98 & 0.05 & 0.24 \\  

\noalign{\vskip 0.1cm}
\hline 
\label{T:Summary_2}
\end{tabular}
\end{table*}


\subsection{The redshift evolution of angular momentum}  

In the context of the $\Lambda$CDM cosmological model, the specific angular momentum of the dark matter halo has a dependency not only with stellar-mass but with time (\citealt{Mo98}). This dependency scales as $j_h\propto M_h^{2/3}(1+z)^{-n}$, with n=1/2 for spherically symmetric haloes in a matter-dominated Universe (\citealt{Obreschkow15}). Assuming that the stellar-to-halo-mass ratio is essentially insensitive to the redshift evolution (\citealt{Behroozi10}), the baryonic component should display a similar behavior $j_*\propto M_*^{2/3}(1+z)^{-1/2}$. These assumptions allow us to investigate the redshift evolution of the specific angular momentum as a function of the environment utilizing the cluster and field samples studied above. In this case, we decided to only use galaxies above $\log M_*=9.5$ for two reasons: First, we showed in Fig.\,\ref{F:AM_OFFSET} that our field samples at different redshifts had a weak to negligible stellar-mass dependence with respect to their specific angular momentum at $\log(M_*)>9.5$. Moreover, the z$\sim$0 sample from \cite{Posti18} hints that in the low stellar-mass regime this dependence may become more important. In addition, our cluster samples are dominated by relatively massive galaxies ($\overline{\log M_*}>10$) and lack objects below $\log M_*=9.5$. 

\begin{figure*}
   \centering
   \includegraphics[width=12cm]{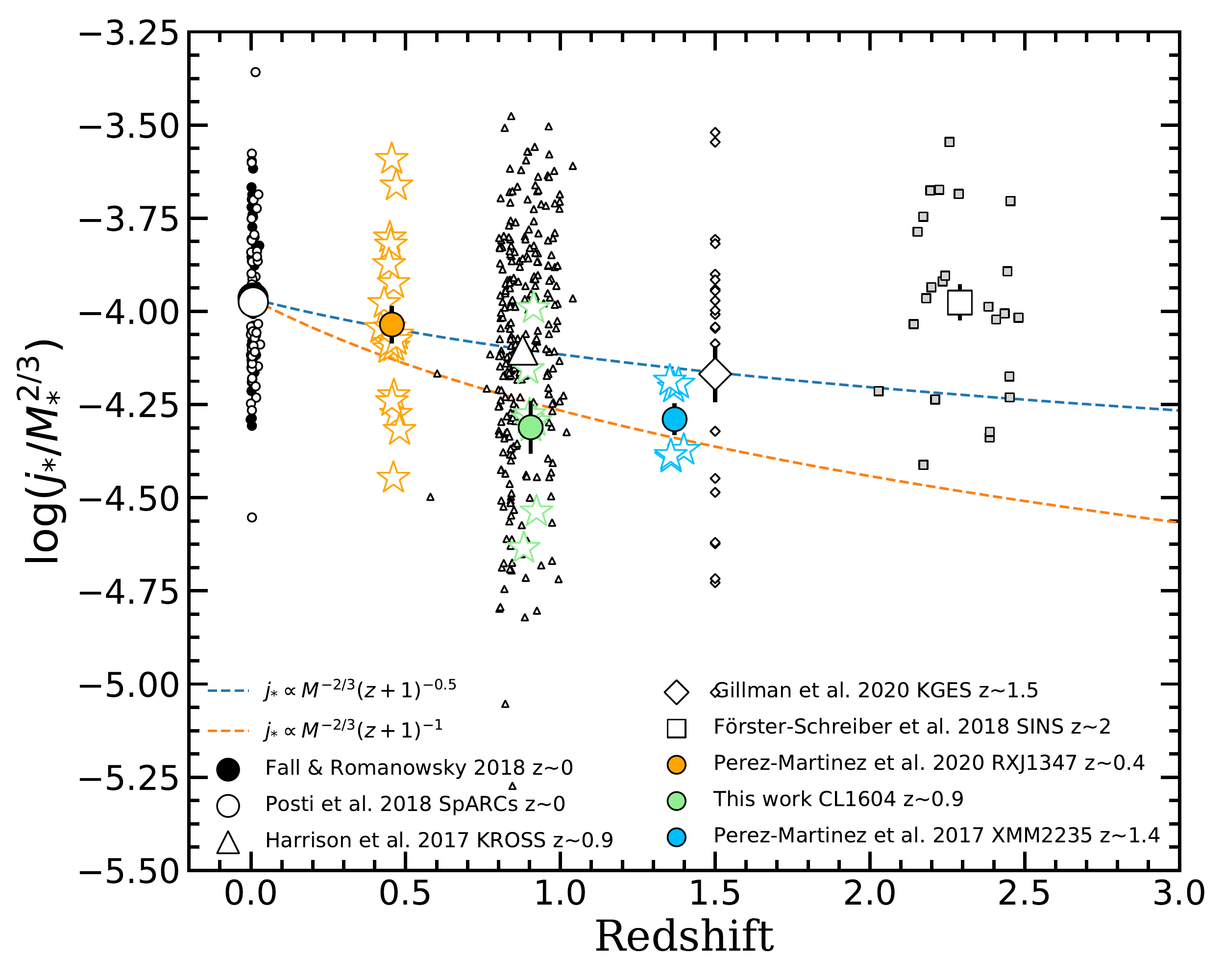}
   \caption{Redshift evolution of $j_*/M^{2/3}$ from z=0 to z$\sim$2.5. The black and empty big circles shows the mean value for the local field galaxies from \citealt{Fall18} (small black cirles) and \citealt{Posti18} (small empty cirles) respectively. The big empty triangle depicts the average value for the KROSS sample at z$\sim$0.9 (small triangles, \citealt{Harrison17}). The empty diamond display the KGES sample at z$\sim$1.5 (small empty diamonds), and the empty square represent the field galaxies at 2<z<2.5 from \citealt{Forster-Schreiber18} (small grey squares). The average $j_*/M^{2/3}$ value for the RXJ1347, CL1604 and XMM2235 cluster samples are respectively shown by the orange, green and blue circles. The errors of the mean are plotted as error bars for each symbol. The dashed lines represent the expected evolution of the angular momentum with redshift according to the $\Lambda CDM$ cosmology: $j_*=M^{-2/3}(z+1)^{-n}$ with n=0.5 (blue) and n=1 (orange).}
   \label{F:ZEVO}
\end{figure*}

After applying this new constraint we present the specific angular momentum redshift evolution in Fig.\,\ref{F:ZEVO}, where the big symbols represent the mean values and their error for each sample included in our angular momentum analysis. We plot the functional form $j_*\propto M_*^{2/3}(1+z)^{-n}$ for n=1/2 and n=1 to allow for different evolutionary paths, and normalize the zero point (z$\sim$0) to the mean values measured from our local Universe local galaxy samples, which are in remarkable agreement: $\log(j_*/M_*^{2/3})=-3.96\pm0.03$ for \cite{Fall18} and $\log(j_*/M_*^{2/3})=-3.97\pm0.03$ for \cite{Posti18}. It is important to emphasize that these two samples are only composed of spiral galaxies, although allowing for small variations in the bulge-to-disc ratios (B/D) to include most late-type galaxies. We find that galaxies at higher redshift (KROSS at z$\sim$0.9 and KGES at z$\sim$1.5) follow the scaling of $j_*\propto M_*^{2/3}(1+z)^{-1/2}$ well, with lower specific angular momentum at higher redshift. In particular,  $\log(j_*/M_*^{2/3})=-4.10\pm0.02$ for the KROSS galaxies and $\log(j_*/M_*^{2/3})=-4.17\pm0.08$ for the KGES survey. These values are equivalent to a specific angular momentum decrease of factor 1.3 by z$\sim$1 in comparison to the local spirals, or a factor 1.6 by z$\sim$1.5, in agreement with the EAGLE numerical simulations (\citealt{Lagos17}). On the other hand, the SINS sample at $2<z<2.5$ is much closer to the local values with $\log(j_*/M_*^{2/3})=-3.98\pm0.05$. A possible explanation for this behavior is the high uncertainties reported for the published structural parameters of this sample (i.e. Sérsic index and $R_e$). 

The trend in the cluster samples, on the other hand, is more difficult to interpret. In general we measure lower $\log(j_*/M_*^{2/3})$ values than in the field, specifically for the higher redshift clusters (e.g. $-4.31\pm0.19$ for CL1604 at z$\sim$0.9, and $-4.29\pm0.10$ for XMM2235 at z$\sim$1.4, but $-4.04\pm0.21$ for RXJ1347 at z$\sim$0.45). Our results suggest that, on average, cluster galaxies have a lower specific angular momentum than their field counterparts at a given epoch. This difference can be explained by the higher probability of interactions that contribute to the angular momentum loss in the cluster environment with respect to the field. Despite not knowing what is the exact contribution of each interaction (e.g. tidal and merging events, ram pressure-stripping, suppression of inflows), these mechanisms appear to be in place as early as z$\sim$1.4 in massive virialized systems such as XMM2235. 



\section{Discussion}
\label{SS:Discussion}

The evolution for nearly isolated galaxies is based on the hierarchical growth of objects and the decline of star formation due to the consumption of gas over cosmic time. However, galaxy clusters strongly influence the evolution of galaxies, introducing cluster-specific interactions that are related to the increasing number density of galaxies towards the central regions of the cluster, the density of the intracluster medium (ICM) and the sub-structures of the cluster. These interactions can be broadly divided into two classes: hydrodynamical, when the gas component acting as a fluid is mainly affected (e.g. ram pressure stripping and starvation), and gravitational when both the stellar and gas component are affected simultaneously by the gravitational potential well of another nearby object (e.g. harassment and mergers). In this section we aim to study the possible contribution of each of these interactions in the evolution of kinematic scaling relations such as the TFR, the VSR, and the AMR.

\subsection{The luminosity and size evolution}

\begin{figure*}
    \begin{multicols}{2}
      \includegraphics[width=\linewidth]{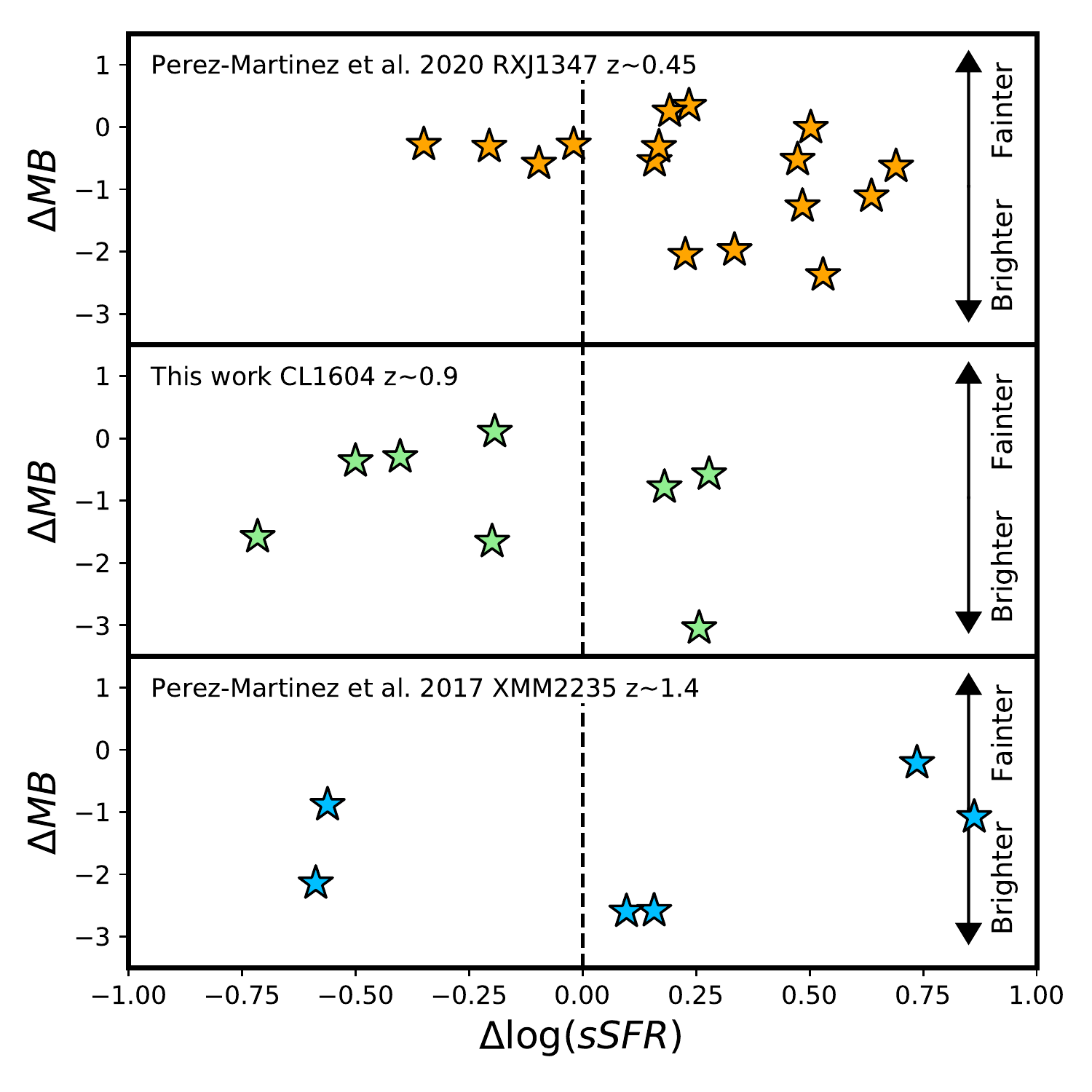}\par
      \includegraphics[width=\linewidth]{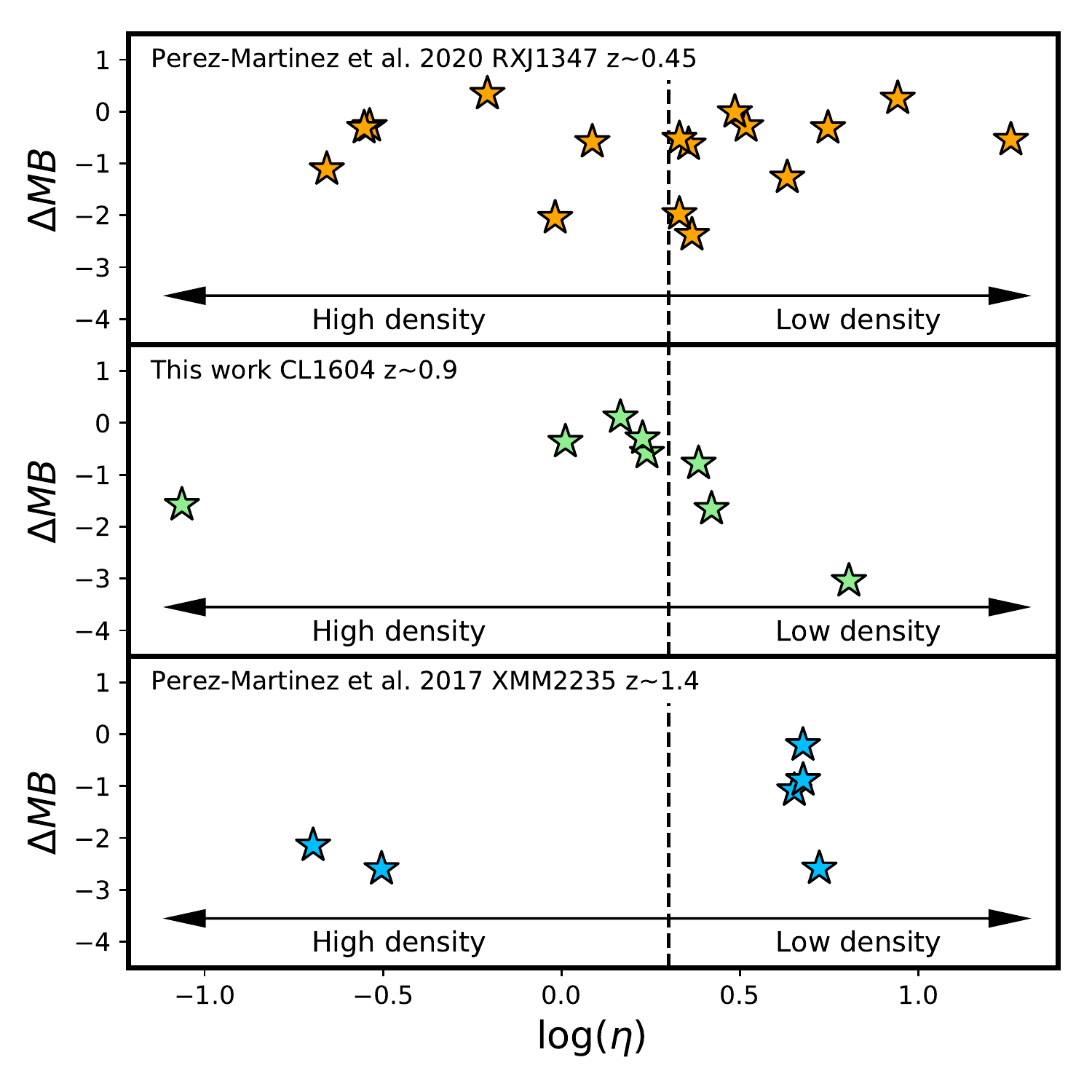}\par
      \end{multicols}
      \caption{Left: B-band TFR offsets as a function of the sSFR offsets measured from the main sequence of star-forming galaxies (\citealt{Peng10}). The color scheme is the same shown in previous figures. Right: B-band offsets with respect to the TFR as a function of global environment ($\eta$) for galaxies in our cluster samples. The dashed lines indicate $\log(\eta)\approx0.3$ which is equivalent to the boundary between accreted and infalling galaxies according to \cite{Noble13}.}
         \label{F:TFR_ENV}
      \end{figure*}

Our work on the B-band TFR (see Fig. \ref{F:TFR}) has shown that cluster galaxies experience a moderate brightening of up to 1 magnitude between $z=0$ and $z=1$. This value is compatible both with the evolution of field galaxies in the same redshift bin (\citealt{Boehm16}) and with the semianalytical model predictions of \cite{Dutton11}, and suggests that the environment is not playing an important role in the B-band luminosity evolution of star-forming galaxies up to this redshift. Therefore, such brightening may be mostly driven by the normal evolution of the galaxies' stellar populations with lookback time. We explore this hypothesis in Fig. \ref{F:TFR_ENV}, where we connect the B-band brightening with the star-formation activity of each object, and with a measurement of the environment they live in through the parameter $\eta$.

In the left-hand panel of Fig. \ref{F:TFR_ENV}, we display the B-band TFR offsets as a function of the sSFR offsets measured from the main sequence of star-forming galaxies discussed in Sect. \ref{SS:sSFR}. We run a Spearman rank correlation test over the cluster samples included in this diagram. Our results show that even though there exists a negative weak correlation between $\Delta M_B$ and $\Delta\log(sSFR)$ for RXJ1347 (r=-0.41) and Cl1604 (r=-0.16), their p-values do not allow to discard the null hypothesis ($p_{RXJ1347}=0.11$ and $p_{Cl1604}=0.69$) which is defined by $p>0.05$. Thus, we conclude that there is no statistical correlation between these parameters for individual cluster galaxies at a fixed redshift. In addition, the right-hand panel of Fig. \ref{F:TFR_ENV} shows no correlation between the luminosity enhancement and the phase-space position of the objects in RXJ1347 and CL1604 as a function of $\eta$. In this diagram $\eta=2$ (i.e. $\log\eta\approx0.3$) divides our galaxy samples in two distinct groups: virialized or recently accreted ($\eta<2$) and infalling ($\eta>2$). The lack of environmental trends between these two groups even when using $\eta$ as a continuous parameter within the accreted region suggests that the influence of the environment on $\Delta M_B$ should be mild in these samples. Similar results were also found when comparing cluster and field galaxies in the B-band TFR at this redshift by \cite{Pelliccia19}. 

There is also no clear correlation between the luminosity enhancements and the sSFR for the XMM2235 objects. However, 2 out of 3 objects with the most negative $\Delta M_B$ values also have very low $\eta$ values and lie within $R_{500}$ of their cluster center (see Fig. 1 in \citealt{JM17}). It has been proposed that the early stages of ram-pressure stripping (RPS) could compress certain regions of the gas kinematics discs causing a temporary enhancement of star formation resulting in a moderate rise of the B-band luminosity (up to $-0.5$ mag) for a given object (\citealt{Ruggiero17,RuggieroThesis}), while the gas kinematics still show relatively ordered rotation (\citealt{Noble19}). This could partially explain the high B-band luminosity values of these two objects for a given $V_{max}$ in the TFR, even though the effects of RPS also depends on the orientation of the disk with respect to the direction of motion within the ICM.

Using $H_{\alpha}$ as SFR estimator, \cite{Boehm20} also found no clear relation between the $\Delta M_B$ and the star-formation activity in the HSC-protoclusters. However, these structures are still in the process of assembling the majority of their mass and building-up their ICM, which makes RPS events unlikely. It has been proposed that during the cluster assembly, galaxies may have higher gas fractions than their field counterparts (\citealt{Noble17}, \citealt{Hayashi18b}). A possible explanation for this behavior is that the filamentary structure of the cosmic web boosts the inflows of pristine gas towards galaxies at its junctions. We speculate that the channeling of fresh cold molecular gas towards the galaxy disk may be responsible for the creation of small starburst that, in turn, enhances the B-band luminosity of the object. However this can not be tested with the optical and NIR data available in these clusters. A simple alternative to the environmentally driven scenarios proposed for XMM2235 and the HSC-protoclusters is that the recipes that govern the star-formation activity in the models of \cite{Dutton11} underestimate the B-band luminosity evolution of TFR objects at high redshift. However the different evolutionary stages of these structures in terms of mass accretion and dynamical state, together with the low number of objects available to study do not allow to draw a decisive conclusion.

 \begin{figure}
      \includegraphics[width=\linewidth]{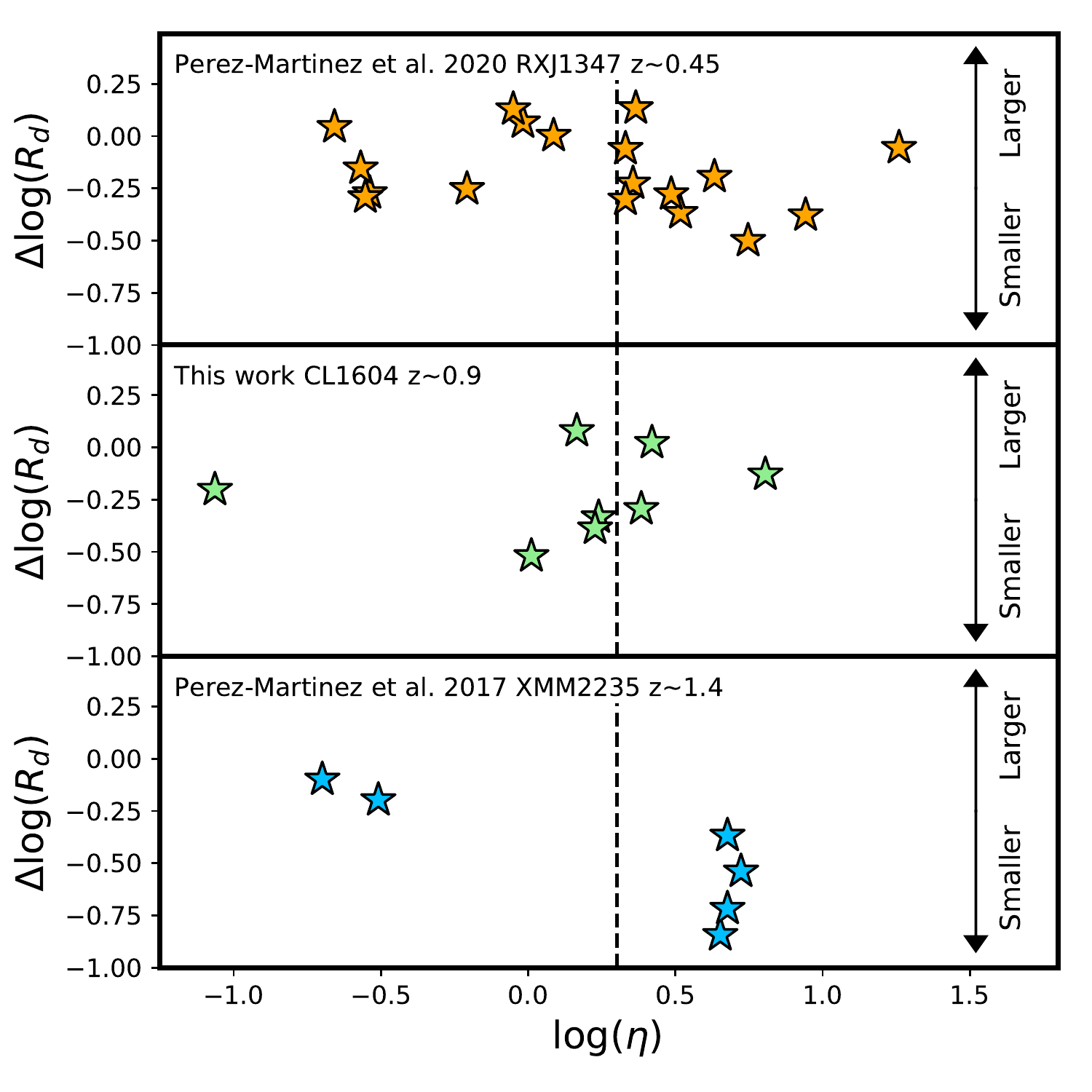}\par
      \caption{Scale length offsets with respect to the VSR as a function of global environment ($\eta$) for galaxies in our cluster samples. The dashed lines indicate $\log(\eta)\approx0.3$ which is equivalent to the boundary between accreted and infalling galaxies according to \cite{Noble13}.}
         \label{F:VFR_ENV}
      \end{figure}
      
We have also tested the influence of the environment on the evolution of the disk size through the VSR. We followed the same approach applied to the TFR by inspecting the offsets of the VSR ($\Delta\log R_d$) with respect to $\eta$. We find no statistically significant trend ($p\gg0.05$) between accreted and infalling galaxies in any of our cluster samples (see Fig. \ref{F:VFR_ENV}). However, we find that the two galaxies that are closer to the innermost regions of XMM2235 show sizes compatible with the local relation. Even though this may not be a general trend due to the few objects involved, it does not support the possibility of these galaxies being ram pressure-stripping candidates, as the truncation of the disc is one of the changes expected in such scenario (\citealt{Crowl08}).

\subsection{The angular momentum evolution} 

Several processes may be responsible for the changes in the specific angular momentum of a galaxy. For example, the outflows of material from the inner parts of a galaxy together with merging events may significantly decrease the specific angular momentum of an object (\citealt{Lagos18a}), while the accretion of cold gas in the outskirts of galaxies residing in cosmic-web filaments (\citealt{Danovich15}) and the migration of clumpy star-forming regions towards the central area of the galaxy may increase it (\citealt{Dekel09b}).

In the cluster environment, where interactions are more frequent than in the field, the loss of angular momentum may become a more important phenomenon. On a first stage, the hot ICM prevents galaxies from getting inflowing material, stopping the build-up of angular momentum that field galaxies experience towards z=0 (\citealt{Peng20}). Furthermore, ram-pressure stripping may also remove the outer parts of the disk gas reservoir when galaxies transit the cluster core causing further angular momentum losses  (\citealt{Romanowsky12}). However, RPS usually requires high ICM density values, which can only be found in the innermost regions of massive clusters. 
We inspect the fraction of retained angular momentum ($f_j$) as a function of the environment in Fig. \ref{F:FJ_ENV}. The number of objects lying in the virialized regions of the cluster (i.e. $\log(\eta)<-0.4$ according to \citealt{Noble13}) is small in all our cluster samples and these objects do not show significantly lower $f_j$ values than objects in the recently accreted ($-0.4<\log(\eta)<0.3$) or infalling ($\log(\eta)>0.3$) regions. Furthermore, RPS does not seem to play a major role in the luminosity and size evolution discussed in the previous section. These two results suggest that RPS can not be taken as the main contributor to the loss of angular momentum in our cluster samples, even though it may be acting on a few individual galaxies. 

 \begin{figure}
      \includegraphics[width=\linewidth]{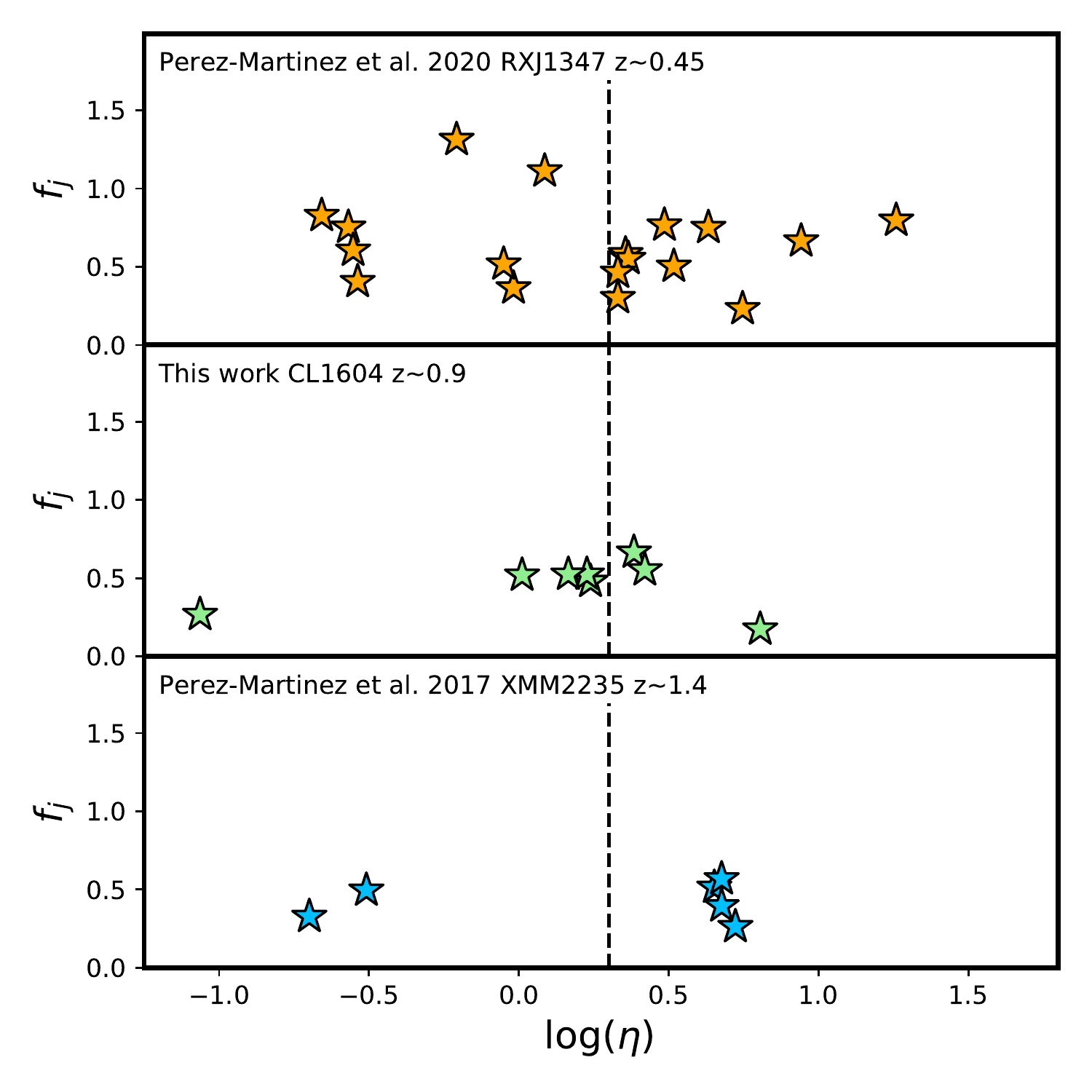}\par
      \caption{Retained angular momentum fraction ($f_j$) as a function of global environment ($\eta$) for galaxies in our cluster samples. The dashed lines indicate $\log(\eta)\approx0.3$ which is equivalent to the boundary between accreted and infalling galaxies according to \cite{Noble13}.}
         \label{F:FJ_ENV}
      \end{figure}

The higher object number density in the cluster environment enhances the frequency of close encounters between galaxies compared to the field (\citealt{Alonso12}), which may also cause angular momentum exchange and loss through tidal interactions and mergers. While the effects of fly-by events might be not as strong as merging events, they may also contribute to increasing the loss of angular momentum in dense environments, especially if repeated over time (\citealt{Moore98}). On the other hand, the destruction of the galaxy disc via major merger in the cluster environment would immediately exclude that kind of galaxy from our analysis. For example, if the major merger is recent, the disturbed gas kinematics prevents those objects from being analyzed in any of the scaling relations discussed in this work, as we could not derive their $V_{max}$. The descendant of a major merging system may be able to re-build a disc at a later stage. However, the time scale involved (several Gyrs, \citealt{Sparre17}) and the need for significant amounts of inflowing gas to build-up the new disc (\citealt{Governato09}, \citealt{Sparre17}) would not allow these systems to carry out this process inside clusters in most cases, especially taking into account that the inflows are cut by the ICM and the typical quenching timescale is $1-2$ Gyrs (\citealt{Wetzel13}, \citealt{Maier19}). 

However, minor mergers may be a possible mechanism for the loss of angular momentum within our sample. Depending on the initial conditions of the event (mass ratio, relative velocity, and geometry), minor mergers could locally affect the stability of the disc without fully destroying its structure (\citealt{Puech12}). \cite{Lagos18a} investigated the influence of mergers on the specific angular momentum finding a great variety of behaviors. In general, major mergers will significantly decrease $j_*$ (on average by a factor 2-3) after 1 Gyr, with counter-rotating dry mergers being especially efficient on this task. The effects of minor wet and dry mergers after 1 Gyr seem to be more subtle, with the former slightly contributing to increase $j_*$ by up to a factor 1.25, especially in the cases when the merger involves high relative orbital velocities and co-rotating geometry, and the latter acting in the opposite direction by decreasing $j_*$ by a similar amount. In both cases, however, the results strongly depend on the exact gas fractions and the geometry of the event. If minor mergers do play a major role in decreasing $j_*$ for cluster galaxies, these interactions should be largely dominated by gas-poor mergers. The average difference between the retained angular momentum fraction of CL1604 and XMM2235 ($f_j\approx0.40$) with respect to the KROSS and the KGES ($f_j\geqslant0.50$) samples is about a $\sim20\%$, which is in principle consistent with the decrease predicted by \cite{Lagos18a} as a consequence of minor mergers. 

Furthermore, mergers are more likely to happen in the outskirts of the clusters than in their cores (\citealt{Deger18}, \citealt{Bahe19}) with minor mergers being also more frequent than major mergers in galaxy groups and cluster outskirts (\citealt{Benavides20}), as it happens in the field (\citealt{Lotz11}, \citealt{Kaviraj15}, \citealt{Ventou19}). The loss of angular momentum through minor mergers in star-forming galaxies before reaching the cluster core could explain why we do not see significant differences in $f_j$ between the population of infalling and accreted cluster regions, as mergers are unlikely to happen in the cluster core and other angular momentum redistribution mechanisms such as RPS may quench these galaxies rather quickly, excluding them from our samples. However, the small number of objects studied per cluster does not allow us to draw a firm conclusion on this regard. Finally, the gas content and relative position of the velocity vectors of merging galaxies are to a large extent stochastic, and thus it is likely that several events of this nature are required to reproduce the observed offsets between the cluster and field samples. We also emphasize that the contribution by other gravitational cluster-specific interactions (e.g. harassment, fly-by encounters, etc) may be as relevant as in the previous case, although it remains unexplored from the numerical simulation point of view to our best knowledge. Larger samples of kinetically analyzed cluster objects are required to shed light onto the role of gravitational interactions as drivers of the angular momentum loss in clusters.

\section{Conclusions} 

In this work, we have studied the redshift evolution of the Tully-Fisher relation, the velocity-size relation, and the angular momentum in clusters up to z$\sim$1.5. We use a collection of clusters at different redshifts for this purpose, with many of them being the single focus of previous publications: Abell 901/902 at z$\sim$0.16 (\citealt{Bosch1, Bosch2}), RXJ1347 at z$\sim$0.45 (\citealt{JM20}), XMM2235 at z$\sim$1.4 (\citealt{JM17}) and two HSC (proto-)clusters at z$\sim$1.5 (\citealt{Boehm20}). Furthermore, we present the first results from a sample of galaxies in the CL1604 cluster system at z$\sim$0.9 observed with GTC/OSIRIS. All cluster samples were studied by our group with similar methods and techniques, which make them ideal for a comparative study between different epochs. We also compare our cluster scaling relations with field samples between z=0 and z=2.5 (\citealt{Boehm16}, \citealt{Harrison17}, \citealt{Fall18}, \citealt{Posti18}, \citealt{Forster-Schreiber18}, \citealt{Gillman20}), cosmological numerical simulations (\citealt{Dutton10, Dutton11}, \citealt{Lagos17,Lagos18a}) and other theoretical works such as \cite{Obreschkow14} and \cite{Obreschkow15}. Our main findings can be summarized as follows:

\begin{enumerate}
    \item Cluster and field galaxies at $0<z<1$ display similar luminosity brightening in the B-band TFR and generally follow the B-band TFR with increasing B-band luminosity values with lookback time. These results are in agreement with the expectations from semianalytical models by \citealt{Dutton11} who predicts $\Delta M_B\approx-1$ mag by $z=1$ due to gradually increasing SFR and younger stellar populations with lookback time. However, we find no correlation between the B-band luminosity enhancement and sSFR within the members of each cluster.
    
    \item Cluster galaxies at z$\sim$1.5 show average B-band luminosity enhancements of $\Delta M_B\approx2$, deviating from the models of \cite{Dutton11} by 0.5-1 mag. This behavior is detected for galaxies residing in massive virialized clusters such as XMM2235 as well as for clusters still in the process of assembling the bulk of their mass such as the two HSC (proto-)clusters studied here. This deviation may be consistent with a moderate underestimation of the modeled luminosity evolution at high redshift. However, based on the conditions of the environment, we can not completely discard the influence of other processes for individual cases. This includes two galaxies near the center of XMM2235 which may be RPS candidates, and the enhancement of SFR  due to higher gas inflows for galaxies in assembling clusters, such as the HSC (proto-)clusters.  
    
    \item Our velocity-size relation results demonstrate that galaxies decrease their disc sizes with redshift at a fixed rotation velocity. Our cluster and field samples follow very similar trends with an average size decrement of factor 1.6 by $z=1$. This result agrees with previous observational constraints (\citealt{Vanderwel14}) and with the numerical models of \cite{Dutton11}. Galaxies in XMM2235 at $z\sim1.4$ are almost 3 times smaller than their local field counterparts, slightly deviating from simulations at the same redshift in the field which predict only a size decrease by a factor of two. However, we find no correlation between the size evolution and $\eta$, which argues against environmental effects as the main drivers for this difference within the limitations of this study in terms of number statistics.
    
    \item Cluster galaxies at $0.5<z<1.5$ follow parallel sequences towards lower $j_*$ values in the "Fall relation" ($j_*\propto M^{2/3}$ \citealt{Fall83, Fall13}) with respect to the local field galaxies. The stellar-to-halo specific angular momentum ratio ($f_j=j_*/j_{halo}$) of cluster galaxies at z$\sim$0.45 is 60$\%$, and drops to less than 40$\%$ by $z\sim0.9$, and remains constant for XMM2235 at z$\sim$1.4. In contrast, field galaxy samples display $\sim20\%$ higher $f_j$ values. This may be caused by cluster-specific interactions. We explore the possible contributions of ram-pressure stripping and mergers based on the kinematic study of our cluster samples (TFR, VSR, and AMR) and suggest that minor mergers may play an important role in angular momentum transformation. 
    
    \item Our analysis of the evolution of specific angular momentum with time yielded different trends for galaxies in the field and cluster environments. The $\Lambda$CDM model predicts a redshift evolution of specific angular momentum in the following form: $j_*\propto M_*^{2/3}(1+z)^{-1/2}$ (\citealt{Mo98}, \citealt{Obreschkow15}). We normalized the zero point of these tracks by re-analyzing the angular momentum contribution of the disc z$\sim$0 galaxies studied in \cite{Fall18} and \cite{Posti18}. The mean values for the field samples analyzed in this study follow the predicted trend up to z$\sim$1.5, although displaying significant scatter (see Fig.\,\ref{F:ZEVO}). However, cluster samples deviate from such a trend at $z\geq1$, falling onto a path better described by $j_*\propto M_*^{2/3}(1+z)^{-1}$. This difference suggests that the effects of the cluster environment over angular momentum are in place as early as z$\sim$1.4 in virialized structures, although further work is needed to pinpoint the specific mechanisms responsible for this difference.
    
\end{enumerate}

The search for environmental effects and their relative importance on the physical parameters of the galaxy populations have been a subject of debate during the last forty years (\citealt{Dressler80}). Scaling relations provide us a way to explore these effects from different perspectives. The upcome of large surveys in the field at intermediate to high redshift, together with new high precision numerical simulations, give us the perfect tools to establish reliable comparisons. However, it is still necessary to increase the number statistics of the cluster samples in the same redshift range and investigate their most important parameters in a comprehensive way (i.e. rotation velocity, size, stellar-mass and populations, metallicity, etc) to fully understand the possible environmental effects at play. The use of multiobject IFU observations and cluster-focused new surveys will be of key importance to disentangle the influence of different cluster-specific interactions over the physical properties of galaxies in the next decade. 

\begin{acknowledgements}

Based on observations made with the Gran Telescopio Canarias (GTC), installed in the Spanish Observatorio del Roque de los Muchachos of the Instituto de Astrof\'isica de Canarias, in the island of La Palma. This paper is also based on data collected at the Subaru Telescope and retrieved from the HSC data archive system, which is operated by Subaru Telescope and Astronomy Data Center at the National Astronomical Observatory of Japan (NAOJ). The HSC Data analysis was in part carried out with the cooperation of Center for Computational Astrophysics, NAOJ. This work is based in part on observations made with the Spitzer Space Telescope, which was operated by the Jet Propulsion Laboratory, California Institute of Technology under a contract with NASA. This work is based in part on observations made with the NASA/ESA Hubble Space Telescope, obtained from the data archive at the Space Telescope Science Institute. STScI is operated by the Association of Universities for Research in Astronomy, Inc. under NASA contract NAS 5-26555. J.M.P. acknowledges the funding support from the University of Vienna and the Marietta Blau Grant, financed by the Austrian Science Ministry, as well as the support from the Instituto de Astrof\'isica de Canarias for allowing him to develop his research in their headquarters at the Canary Islands, Spain. H.D. acknowledges financial support from the Spanish Ministry of Science, Innovation and Universities (MICIU) under the 2014 Ramón y Cajal program RYC-2014-15686 and AYA2017-84061-P, the latter one co-financed by FEDER (European Regional Development Funds). This work has been also supported by Direcci\'on General de Investigaci\'on Cient\'ifica y T\'ecnica (DGICYT) grant AYA2016-79724-C4-1-P.
          
\end{acknowledgements}

\bibliographystyle{aa.bst} 
\bibliography{references.bib} 

\begin{appendix}
\section{Additional material}
In this section, we present the data tables containing all the relevant parameters of the RXJ1347, CL1604 and XMM2235 cluster sample. In addition, we display the observed and synthetic rotation curves for the Cl1604 objects.
\label{A:app}

\begin{sidewaystable*}[h!]
\centering
\caption{General properties of the Cl1604 galaxy sample. IDs, J2000 coordinates, redshift, dust extinction corrected B-band absolute magnitudes in the AB system, logarithmic stellar-mass, maximum rotation velocity and its error, effective radius, inclination, misalignment angle, SED reddening, [OII] star-formation rate, environmental parameter and its error.}
\begin{tabular}{ccccccccccccccc}
\hline
\noalign{\vskip 0.1cm}
ID & RA & DEC & $z$ & $M_{B'}$ & $\log{M_{\ast}}$ & $V_{max}$ & $V_{max,err}$ & $R_{e}$ & $i$ & $\delta$ & E(B-V) & $SFR_{[OII]}$ & $\eta$ & $\eta_{err}$\\
   &  hh:mm:ss.s & dd:mm:ss.s &   & (mag)    &   &  (km/s)   & (km/s)   & (kpc)  & (deg)   & (deg) &    & ($M_{\odot}/yr$) & \\
\noalign{\vskip 0.1cm}
\hline 
\hline 
\noalign{\vskip 0.2cm}
A1 & 16:04:24.7 & 43:18:35.9 & 0.9228 & -21.32 & 10.63 & 145 & 17 & 3.8 & 62 & 9  & 0.2 & 4.2  & 0.09 & 0.02 \\
A2 & 16:04:30.0 & 43:19:59.4 & 0.9183 & -21.62 & 10.65 & 213 & 19 & 4.1 & 63 & 3  & 0.5 & 42.6 & 1.74 & 0.44 \\
A3 & 16:04:23.0 & 43:02:07.4 & 0.8945 & -21.13 & 10.27 & 134 & 16 & 4.9 & 62 & 18 & 0   & 7.3  & 2.63 & 1.06 \\
A4 & 16:04:19.0 & 43:03:28.9 & 0.8815 & -21.44 & 10.28 & 97  & 16 & 2.6 & 61 & 4  & 0.3 & 21.2 & 6.39 & 2.57 \\
A5 & 16:04:25.8 & 43:05:10.0 & 0.9132 & -21.22 & 9.99  & 178 & 14 & 4.8 & 33 & 19 & 0.1 & 10.7  & 2.42 & 0.98 \\
A6 & 16:04:18.8 & 43:05:50.2 & 0.9047 & -22.22 & 10.99 & 276 & 22 & 6.3 & 56 & 4 & 0.2 & 16.1  & 1.69 & 0.68 \\
A7 & 16:04:18.0 & 43:06:03.1 & 0.8953 & -22.47 & 10.85 & 291 & 21 & 8.2 & 73 & 16 & 0.1 & 10.1  & 1.03 & 0.41 \\
A8 & 16:04:32.1 & 43:07:12.2 & 0.9005 & -19.73 & 10.70 & 149 & 14 & 6.0 & 46 & 14 & 0.5 & 15.7  & 1.46 & 0.59 \\

\noalign{\vskip 0.2cm}
\hline 
\noalign{\vskip 0.2cm}
\label{T:CL16}
\end{tabular}
\end{sidewaystable*}

\begin{sidewaystable*}[h!]
\centering
\caption{General properties of the RXJ1347 galaxy sample. IDs, J2000 coordinates, redshift, dust extinction corrected B-band absolute magnitudes in the AB system, logarithmic stellar-mass, maximum rotation velocity and its error, effective radius, inclination, misalignment angle, SED reddening, [OII] star-formation rate, environmental parameter and its error.} 
\begin{tabular}{ccccccccccccccc}
\hline
\noalign{\vskip 0.1cm}
ID & RA & DEC & $z$ & $M_{B'}$ & $\log{M_{\ast}}$ & $V_{max}$ & $V_{max,err}$ & $R_{e}$ & $i$ & $\delta$ & E(B-V) & $SFR_{[OII]}$ & $\eta$ & $\eta_{err}$\\
   &  hh:mm:ss.s & dd:mm:ss.s &   & (mag)    &   &  (km/s)   & (km/s)   & (kpc)  & (deg)   & (deg) &    & ($M_{\odot}/yr$) & \\
\noalign{\vskip 0.1cm}
\hline 
\hline 
\noalign{\vskip 0.2cm}
B1 & 13:48:29.2 & -11:34:39.1 & 0.4195 & -22.02 & 10.93 & 250  & 8  & 10.9 & 38.3 & 28.1 & 0.5 & 24.2  & 18.16 & 3.48 \\
B2 & 13:48:13.6 & -11:37:12.3 & 0.4457 & -21.29 & 10.46 & 205  & 12 & 5.5  & 78.0 & 6.4  & 0.5 & 36.1 & 2.27 & 0.43 \\
B3 & 13:48:17.9 & -11:38:27.4 & 0.4609 & -21.80 & 10.18 & 196  & 4  & 5.6  & 85.6 & 1.4  & 0.1 & 13.8 & 4.29 & 0.82 \\
B4 & 13:47:36.9 & -11:36:07.5 & 0.4551 & -20.17 & 9.61  & 129  & 8  & 5.8  & 39.7 & 1.3  & 0   & 1.3  & 1.22 & 0.23 \\
B5 & 13:47:47.0 & -11:37:31.7 & 0.4488 & -22.26 & 10.73 & 337  & 14 & 10.4 & 45.4 & 0.7  & 0.4 & 20.3  & 0.62 & 0.12 \\
B6 & 13:47:28.9 & -11:38:12.7 & 0.4700 & -20.67 & 9.58  & 138  & 4  & 5.8  & 49.5 & 2.3  & 0   & -    & 0.02 & 0.01\\
B7 & 13:47:39.3 & -11:39:20.7 & 0.4521 & -21.04 & 9.95  & 148  & 3  & 6.7  & 64.0 & 1.9  & 0.3 & 13.1  & 0.22 & 0.04 \\
B8 & 13:47:04.2 & -11:51:50.2 & 0.4619 & -21.65 & 10.80 & 259  & 12 & 5.5  & 45.9 & 1.6  & 0.3 & 6.0  & 3.29 & 0.63 \\
B9 & 13:47:09.9 & -11:57:22.9 & 0.4306 & -21.05 & 10.20 & 225  & 8  & 4.4  & 44.9 & 12.1 & 0.3 & 7.3  & 8.75 & 1.68 \\
B10 & 13:46:34.7 & -11:51:30.2 & 0.4615 & -21.66 & 10.32 & 132 & 6  & 7.0 & 76.8  & 6.5  & 0.2 & 19.5 & 2.32 & 0.68 \\
B11 & 13:46:19.9 & -11:53:00.0 & 0.4683 & -20.92 & 10.55 & 191 & 19 & 4.5 & 64.8  & 0.1  & 0.5 & 8.2  & 0.29 & 0.09 \\
B12 & 13:46:30.8 & -11:53:43.0 & 0.4802 & -21.19 & 10.52 & 129 & 10 & 5.8 & 53.2  & 8.9  & 0.2 & 13.8  & 0.96 & 0.28  \\
B13 & 13:46:30.6 & -11:53:55.8 & 0.4733 & -21.82 & 10.85 & 256 & 10 & 8.9 & 72.4  & 15.3 & 0.3 & -    & 0.27 & 0.08 \\
B14 & 13:46:32.3 & -11:55:49.9 & 0.4555 & -20.64 & 9.91  & 105 & 4  & 3.2 & 71.8  & 5.1  & 0.1 & 6.1  & 2.14 & 0.63 \\
B15 & 13:46:28.3 & -11:56:52.7 & 0.4461 & -20.38 & 10.03 & 161 & 19 & 6.4 & 50.6  & 4.6  & 0.4 & 11.1  & 3.06 & 0.59 \\
B16 & 13:46:37.3 & -11:57:16.6 & 0.4473 & -20.37 & 10.02 & 167 & 7  & 5.3 & 71.0  & 3.4  & 0.3 & 10.2  & 2.14 & 0.41 \\
B17 & 13:46:51.6 & -11:47:24.1 & 0.4482 & -20.71 & 10.02 & 115 & 11 & 5.7 & 54.3  & 21.7 & 0.3 & -    & 0.89 & 0.17 \\
B18 & 13:46:16.2 & -11:47:01.6 & 0.4604 & -20.52 & 10.26 & 166 & 5  & 2.1 & 38.3  & 18.4 & 0.2 & 7.7  & 5.58 & 1.07 \\
B19 & 13:46:36.1 & -11:47:46.7 & 0.4695 & -21.21 & 10.43 & 219 & 6  & 5.1 & 60.9  & 13.0 & 0.2 & 4.4  & 0.28 & 0.08 \\

\noalign{\vskip 0.2cm}
\hline 
\noalign{\vskip 0.2cm}
\label{T:RXJ}
\end{tabular}
\end{sidewaystable*}

\begin{sidewaystable*}[h!]
\centering
\caption{General properties of the XMM2235 galaxy sample. IDs, J2000 coordinates, redshift, dust extinction corrected B-band absolute magnitudes in the AB system, logarithmic stellar-mass, maximum rotation velocity and its error, effective radius, inclination, misalignment angle, SED reddening, [OII] star-formation rate, environmental parameter and its error.} 
\begin{tabular}{ccccccccccccccc}
\hline
\noalign{\vskip 0.1cm}
ID & RA & DEC & $z$ & $M_{B'}$ & $\log{M_{\ast}}$ & $V_{max}$ & $V_{max,err}$ & $R_{e}$ & $i$ & $\delta$ & E(B-V) & $SFR_{[OII]}$ & $\eta$ & $\eta_{err}$\\
   &  hh:mm:ss.s & dd:mm:ss.s &   & (mag)    &   &  (km/s)   & (km/s)   & (kpc)  & (deg)   & (deg) &    & ($M_{\odot}/yr$) & \\
\noalign{\vskip 0.1cm}
\hline 
\hline 
\noalign{\vskip 0.2cm}
C1 & 22:35:33.5 & -25:55:08.7 & 1.3658 & -23.71 & 10.48 & 340 & 24 & 2.3 & 65 & 5  & 0.4 & 245.7 & 4.49 & 0.75 \\
C2 & 22:35:33.1 & -25:55:47.1 & 1.3576 & -22.58 & 10.03 & 156 & 19 & 1.7 & 75 & 4  & 0.3 & 22  & 5.28 & 0.88 \\
C3 & 22:35:21.8 & -25:57:14.0 & 1.3986 & -21.52 & 10.37 & 130 & 12 & 3.7 & 70 & 10 & 0.1 & 7.2   & 0.20 & 0.03 \\
C4 & 22:35:18.1 & -25:58:06.6 & 1.3816 & -22.96 & 10.40 & 174 & 11 & 4.2 & 76 & 4  & 0   & 36.6  & 0.31 & 0.05 \\
C5 & 22:35:27.8 & -25:59:48.8 & 1.3534 & -22.46 & 10.68 & 249 & 35 & 4.7 & 44 & 26 & 0   & 13.1  & 4.75 & 0.79 \\
C6 & 22:35:17.1 & -26:00:26.3 & 1.3565 & -22.78 & 10.91 & 334 & 34 & 3.2 & 34 & 14 & 0.4 & 390 & 4.75 & 0.79 \\

\noalign{\vskip 0.2cm}
\hline 
\noalign{\vskip 0.2cm}
\label{T:XMM}
\end{tabular}
\end{sidewaystable*}

\begin{figure*}[h!]
\centering
\includegraphics[width=\textwidth]{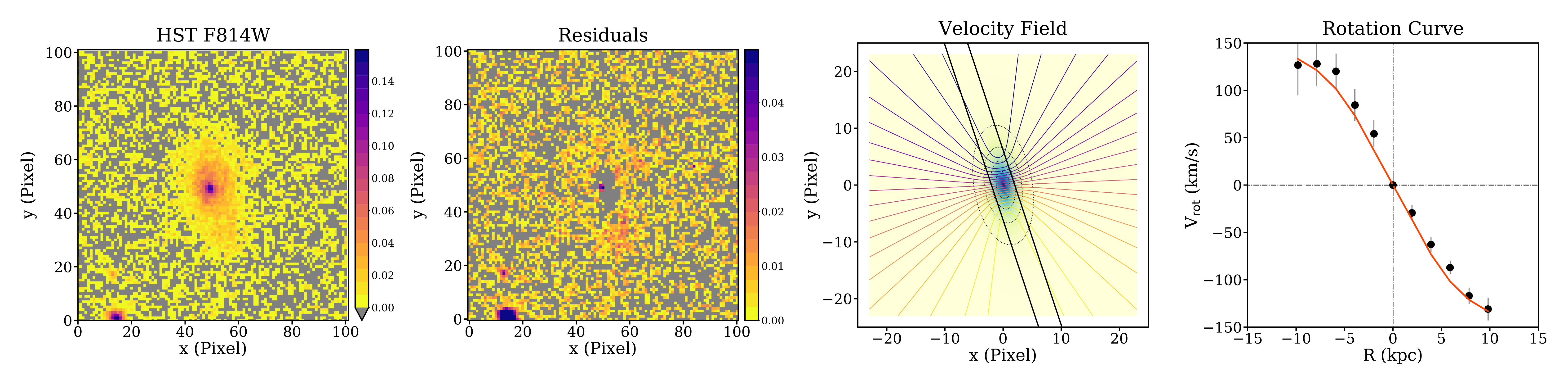}
\includegraphics[width=\textwidth]{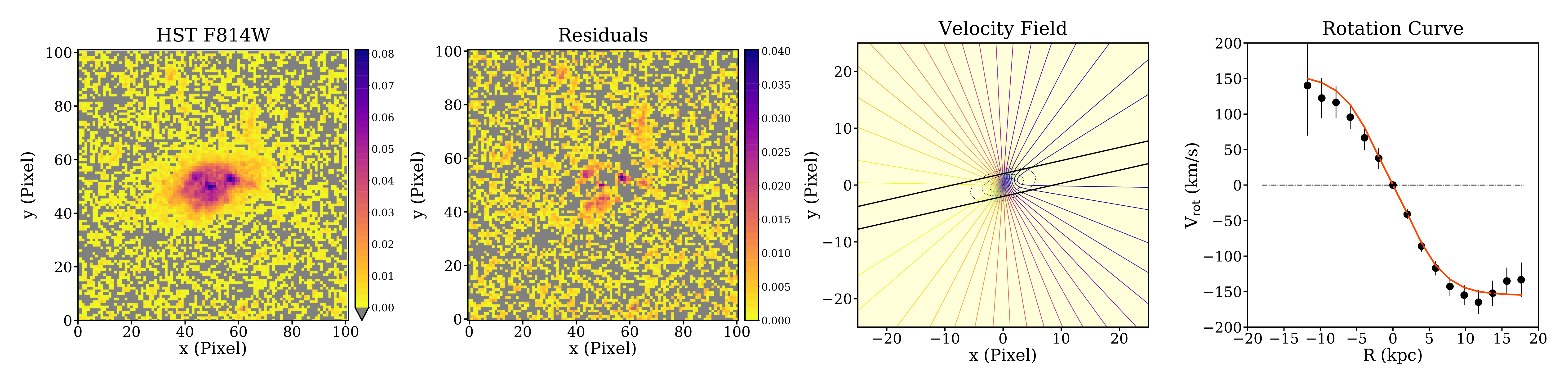}
\includegraphics[width=\textwidth]{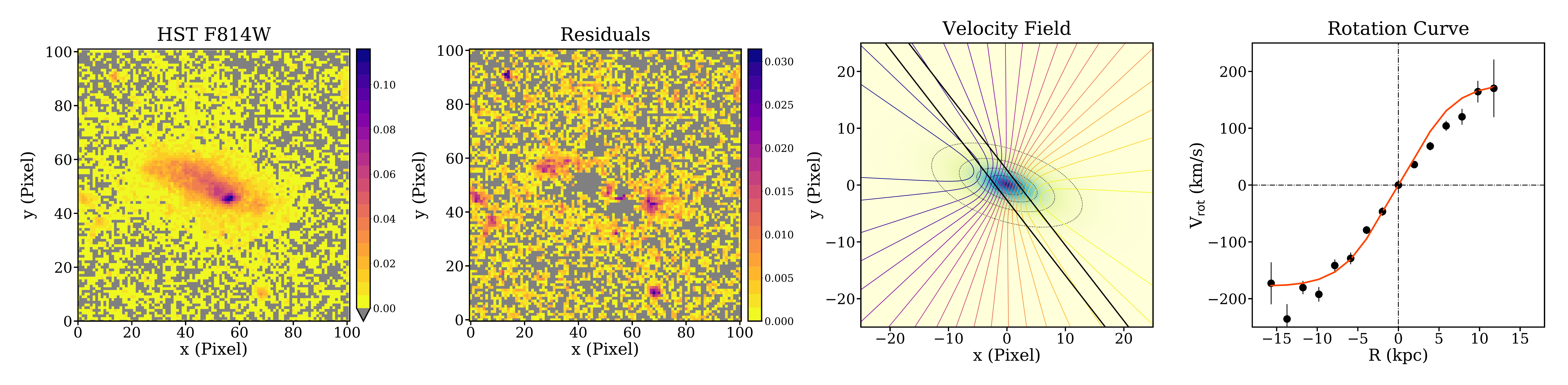}
\includegraphics[width=\textwidth]{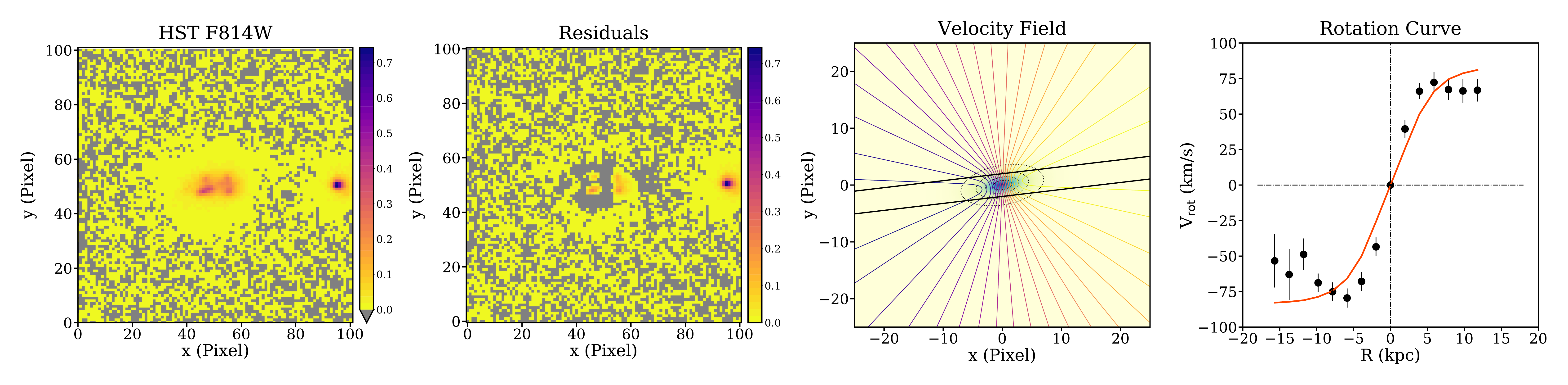}
\includegraphics[width=\textwidth]{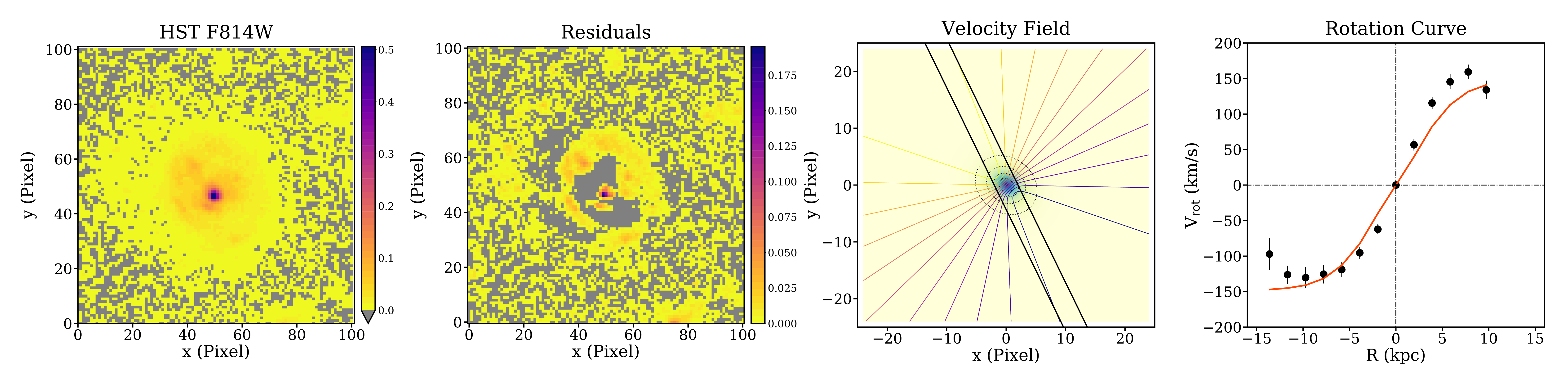}
\caption{The CL1604 sample of galaxies introduced in Sect. \ref{SS:Methods} and presented in the same order as in Table \ref{T:CL16}. The first and second columns respectively show the original HST-F814W or HSC-z-band images centered on the targets and their residuals after subtracting the 2D model of the galaxies. Note that the pixel scale in the first column corresponds to 0.05"/pix for HST images and 0.2"/pix for HSC images. The third column presents the synthetic velocity field after fitting the simulated rotation curve to the observed curve (assuming the pixel scale of OSIRIS, i.e. 0.25"/pix). The black solid parallel lines depict the edges of the slit. The fourth column displays the observed (black dots) and modelled (red line) rotation curve. }
\ContinuedFloat
\label{foot}
\end{figure*}

\begin{figure*}[h!]
\centering
\includegraphics[width=\textwidth]{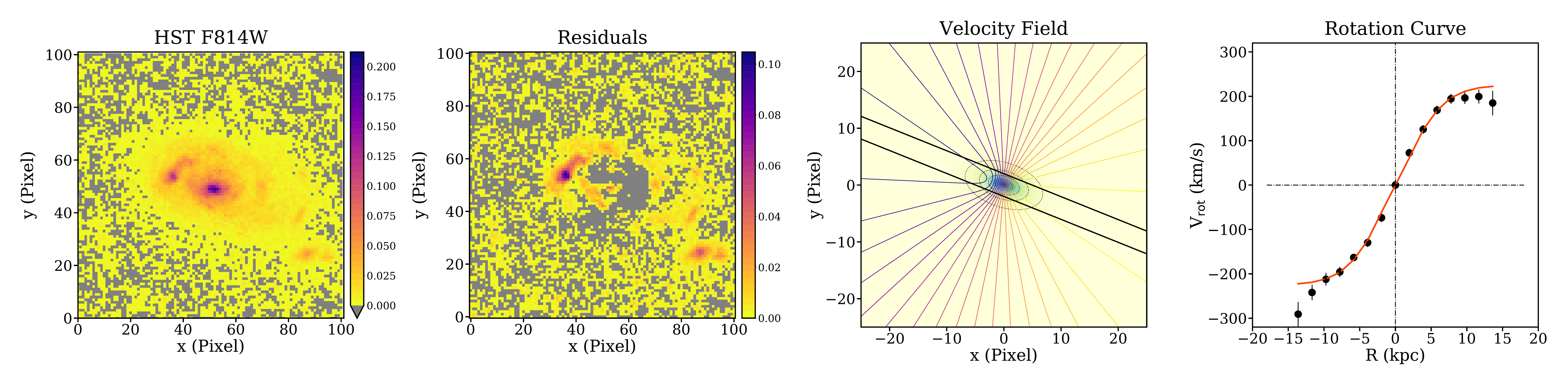}
\includegraphics[width=\textwidth]{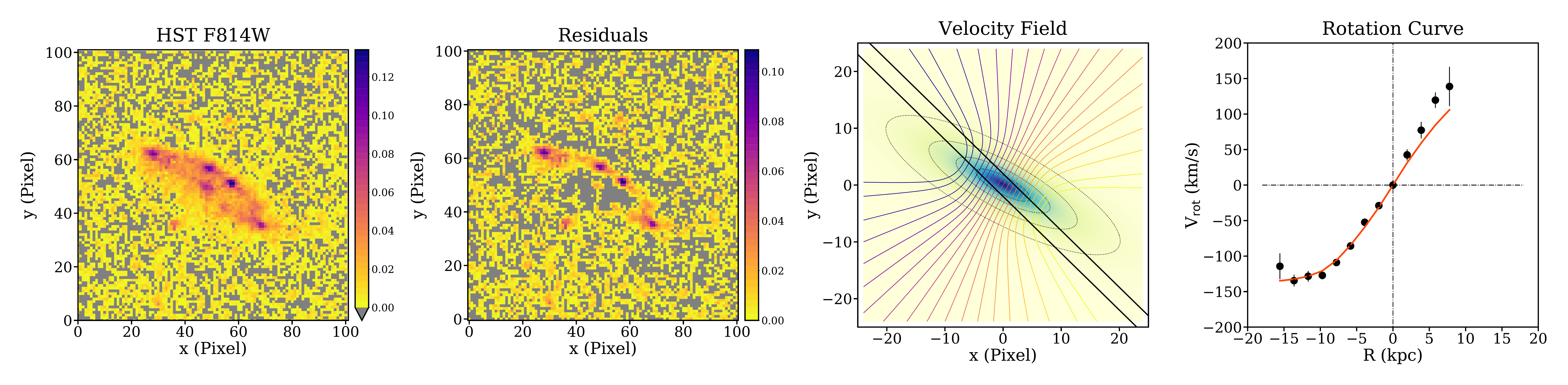}
\includegraphics[width=\textwidth]{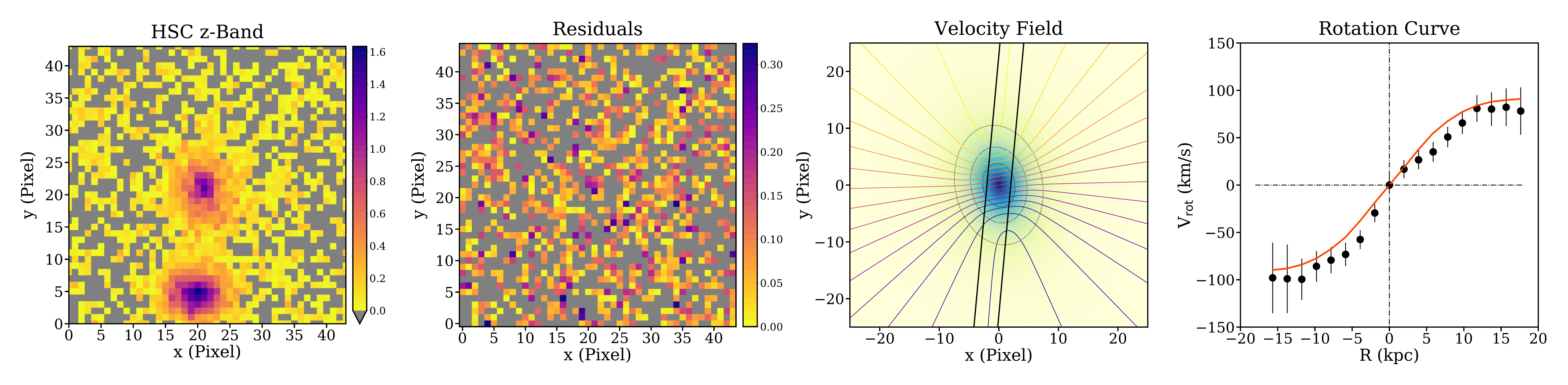}

\caption[]{(Continued)}
\ContinuedFloat
\label{foot}
\end{figure*}    

\end{appendix}
\end{document}